%% file: main.tex
\documentclass{article}
\usepackage[a4paper,top=2cm,bottom=2cm,left=3cm,right=3cm,marginparwidth=1.75cm]{geometry}
\usepackage{amsmath,amssymb,amsfonts}
\usepackage{graphicx}
\usepackage[dvipsnames]{xcolor}
\usepackage{amsthm}
\usepackage{xspace}
\usepackage{tabularx,booktabs}
\usepackage{multirow}
\usepackage{changepage}
\usepackage{nicefrac}
\usepackage{tikz}
\usetikzlibrary{shapes, snakes}
\usetikzlibrary{backgrounds}
\usepackage[many]{tcolorbox}
\usepackage{array}
\usepackage[natbib=true,sortcites,maxbibnames=50]{biblatex}
\usepackage[pdfdisplaydoctitle,menucolor=orange!40!black,filecolor=magenta!40!black,urlcolor=blue!40!black,linkcolor=red!40!black,citecolor=green!40!black,colorlinks]{hyperref}
\usepackage[nameinlink,capitalize]{cleveref}
\crefname{figure}{Figure}{Figures}
\crefname{rrule}{Reduction Rule}{Reduction Rules}
\usepackage{todonotes}
\usepackage{authblk}

\newtheorem{theorem}{Theorem}[section]
\newtheorem{lemma}[theorem]{Lemma}
\newtheorem{corollary}[theorem]{Corollary}

\newtheorem{observation}[theorem]{Observation}

\theoremstyle{definition}
\newtheorem{definition}{Definition}[section]
\newtheorem{rrule}[theorem]{Reduction Rule}

\newcommand{\sr}{\textsc{Segment Routing}\xspace}
\newcommand{\unitsr}{\textsc{Unit Segment Routing}\xspace}
\newcommand{\unarysr}{\textsc{Unary Segment Routing}\xspace}
\newcommand{\threeedgecol}{\textsc{3-Edge Coloring}\xspace}

\newcommand{\binpacking}{\textsc{Bin Packing}\xspace}
\newcommand{\unarybinpacking}{\textsc{Unary Bin Packing}\xspace}

\newcommand{\twodoneplong}{\textsc{2 Disjoint 1 Shortest Path}\xspace}
\newcommand{\twodonep}{\textsc{2D1SP}\xspace}
\newcommand{\twoedplong}{\textsc{2 Edge Disjoint Paths}\xspace}
\newcommand{\twoedp}{\textsc{2-EDP}\xspace}
\newcommand{\mcc}{\textsc{Multicolored Clique}\xspace}
\newcommand{\threepart}{\textsc{3 Partition}\xspace}
\newcommand{\lsr}{\langle}
\newcommand{\rsr}{\rangle}

\tcbset{
    sharp corners,
    colback = white,
    before skip = 0.4cm,
    after skip = 0.4cm
}
\newtcolorbox{boxA}{
    colframe = black,
    boxrule = 0.5pt,
    right = 3pt,
    left = 3pt,
    top = 3pt,
    bottom = 3pt,
}

\newcommand{\pbdef}[3]{
\begin{boxA}
    \begin{tabular}{@{}l p{12.25cm}}
    \multicolumn{2}{@{}l}{#1} \\
    Input:      & #2 \\
    Question:   & #3 \\
\end{tabular}
\end{boxA}
}

\tikzset{
    black vertex/.style={circle, fill=black, inner sep=0pt, minimum size=5pt},
    gray vertex/.style={circle, fill=gray, inner sep=0pt, minimum size=5pt},
    small black vertex/.style={circle, fill=black, inner sep=0pt, minimum size=4pt},
    very small black vertex/.style={circle, fill=black, inner sep=0pt, minimum size=3pt},
    small gray vertex/.style={circle, fill=gray, inner sep=0pt, minimum size=4pt},
    very small gray vertex/.style={circle, fill=gray, inner sep=0pt, minimum size=2.5pt},
    round/.style={circle, fill=red, inner sep=0pt, minimum size=5pt},
    square/.style={rectangle, fill=RoyalBlue, inner sep=0pt, minimum size=5pt},
    lozenge/.style={diamond, fill=Green, inner sep=0pt, minimum size=7pt},
    triangle/.style={regular polygon,regular polygon sides=3, fill=darkgray, inner sep=0pt, minimum size=7pt},
    square waypoint/.style={draw=black, fill=none, inner sep=0pt, minimum size=8pt, thick},
    round waypoint/.style={circle, draw=black, fill=none, inner sep=0pt, minimum size=8pt, thick},
    gray edge/.style={color=gray, very thin},
    black edge/.style={color=black, thick},
    gray arc/.style={->, color=gray, very thin},
    black arc/.style={->, color=black, thick},
    dashed edge/.style={color=black, dash pattern=on 5pt off 4pt, very thick},
    dotted edge/.style={color=black, dotted, very thick},
    dashdotted edge/.style={color=black, dash pattern=on 4pt off 3pt on 1.5pt off 3pt, very thick},
    dashed arc/.style={->, color=black, dash pattern=on 5pt off 4pt, very thick},
    dotted arc/.style={->, color=black, dotted, very thick},
    dashdotted arc/.style={->, color=black, dash pattern=on 4pt off 3pt on 1.5pt off 3pt, very thick},
}

\begin{document}

\title{Parameterized Complexity of Segment Routing}
\author[1]{Cristina Bazgan}
\author[2]{Morgan Chopin}
\author[3]{André Nichterlein}
\author[1,2]{Camille Richer}
\affil[1]{\normalsize Université Paris-Dauphine, PSL Research University, CNRS, UMR 7243, LAMSADE, Paris, France}
\affil[2]{\normalsize Orange Innovation, Châtillon, France}
\affil[3]{\normalsize Algorithmics and Computational Complexity, Technische Universität Berlin, Germany}
\affil[ ]{cristina.bazgan@dauphine.fr, morgan.chopin@orange.com, andre.nichterlein@tu-berlin.de, camille.richer@orange.com}
\date{}

\maketitle

\begin{abstract}
	Segment Routing is a recent network technology that helps optimizing network throughput by providing finer control over the routing paths. 
	Instead of routing directly from a source to a target, packets are routed via intermediate waypoints. 
	Between consecutive waypoints, the packets are routed according to traditional shortest path routing protocols. 
	Bottlenecks in the network can be avoided by such rerouting, preventing overloading parts of the network.
	The associated NP-hard computational problem is \sr{}: Given a network on~$n$ vertices, $d$~traffic demands (vertex pairs), and a (small) number~$k$, the task is to find for each demand pair at most~$k$ waypoints such that with shortest path routing along these waypoints, all demands are fulfilled without exceeding the capacities of the network.
	We investigate if special structures of real-world communication networks could be exploited algorithmically.
	Our results comprise NP-hardness on graphs with constant treewidth even if only one waypoint per demand is allowed.
	We further exclude (under standard complexity assumptions) algorithms with running time~$f(d) n^{g(k)}$ for any functions~$f$ and~$g$.
	We complement these lower bounds with polynomial-time solvable special cases.
\end{abstract}

\section{Introduction}

With the arrival of next-generation networks enabled by disruptive technologies such as Software-Defined Networking (SDN) and Network Virtualization, network managers face several unprecedented challenges. These include the adaptation of networks to accommodate the ever-growing volume of traffic and the multiplicity of user services and content, all while maintaining high quality of service (QoS).
This situation highlights the need for scalable and efficient traffic engineering techniques to optimize the utilization of network resources. 
Despite the potential of more flexible protocols like Multi Protocol Label Switching (MPLS), most IP networks still rely primarily on shortest path routing protocols such as Open Shortest Path First (OSPF)~\cite{rfc2328} or Intermediate System to Intermediate System (IS-IS)~\cite{rfc1142}. 
Although these protocols are easy to manage, they offer limited control over the induced routing paths, as adjustments can only be made indirectly by changing the network link weights. 
Additionally, network managers prefer solutions that require minimal changes to their network configuration.

\begin{figure}[t]
    \centering
    \begin{tikzpicture}[xscale=1.2]
		\footnotesize
        \input{figures/intro-example}
    \end{tikzpicture}
    \caption{On the left, the data are routed from~$a$ to~$g$ with traditional shortest path routing. The fraction on each arc indicates how the data split among the multiple shortest paths. On the right, a waypoint in~$c$ is introduced, resulting in the data being routed through the bottom part of the network and freeing up capacities in the top part.}
    \label{fig:definitions}
\end{figure}
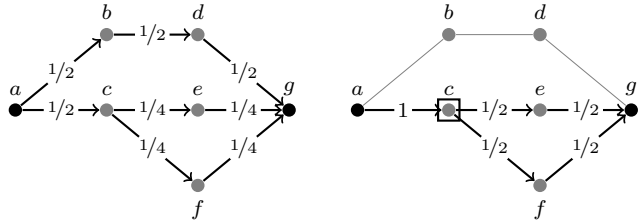

In this context, Segment Routing (SR)~\cite{rfc8402} appears as a promising approach, enabling the possibility to route packets over non-shortest paths without extensive modifications to the network. 
Specifically, each packet entering the network is assigned a so-called \emph{segment path}, which is a sequence of routers referred to as \emph{node segments} or \emph{waypoints} that the packet must follow in the network. Some works also consider \emph{adjacency segments}, which are links of the network. Between two segments, traditional shortest path-based routing is used. 
By encoding routing instructions directly into the packet header, SR allows packets to deviate from the shortest paths, offering greater flexibility with minimal additional implementation costs for network administrators (see \cref{fig:definitions}).
Hence SR is of great practical interest for traffic engineering in networks based on shortest path routing protocols. 
We refer to \citet{AFTU13} for a general survey on variants of SR.

In this paper, we are interested in solving the corresponding decision problem \sr{}:
We are given an (un)directed graph~$G$ on $n$ vertices and $m$ arcs, whose arcs have capacities and weights, a set of $d$ demands (or commodities) and an integer~$k$.
The task is to assign a sequence of at most~$k$ waypoints to each demand such that the flow routed on the shortest paths between consecutive waypoints is feasible, \emph{i.\,e.}, the total demand volume traversing an arc is not greater than its capacity.
Observe that an algorithm solving \sr{} could be combined with binary search to find segment paths that minimize the maximum congestion over all arcs, also called \emph{maximum link utilization}~(MLU).\footnote{Perform binary search over~$x \in [0,100]$ where at most $x\,\%$ of every edge capacity can be used by the algorithm.}

Approaches based on linear programming~\cite{bhatia2015optimized,jadin2019cg4sr,callebaut2023preprocessing} or constraint programming \cite{hartert2015solving} have been employed in previous works to solve optimization versions of \sr{}. Recent works have also studied ways to implement segment routing as traffic engineering tunnels instead of configuring one segment path per demand \cite{brundiers2022midpoint}.
We complement these practical approaches with a rigorous analysis of the computational complexity of \sr{}, a variant of which is known to be weakly NP-hard~\cite{hartert2015solving}. Such analysis has been conducted for the related problem \textsc{Waypoint Routing} \cite{AFJS18, AFJPS18, AFS20, SS22}, in which the task is to find a (shortest) walk visiting a given set of waypoints and respecting edge capacities.

\begin{table}[t]
    \centering
    \caption{
		Summary of the results and of the used parameters (in top part all numbers in the input are one, in the middle part all numbers are encoded in unary).
	}
    \begin{tabularx}{0.95\linewidth}{llll}
		\toprule
		& param. & undirected & directed \\
		\midrule
		\multirow{4}{*}{unit} & $k$         & NP-hard if~$k=1$ (\cref{thm:threeedgecol}) & NP-hard if~$k=1$ (\cref{thm:mcc}) \\
		& $k+\tau$    & NP-hard if~$k=1, \tau = 4$ (\cref{thm:threeedgecol}) & ? \\
		& $d$         & NP-hard if~$d=4$ (\cref{thm:twodonep}) & NP-hard if~$d=3$ (\cref{thm:twoedp}) \\
		& $d+k$       & \multicolumn{2}{c}{W[1]-hard (\cref{thm:mcc})} \\ 
		\midrule
		\multirow{2}{*}{unary} & $n$  &  \multicolumn{2}{c}{W[1]-hard even if~$k=1$ (\cref{cor:binpacking})} \\ 
		& $k+\tau$ & \multicolumn{2}{c}{NP-hard if~$k=1, \tau = 2$ (\cref{thm:binpacking})} \\
		\bottomrule
    \end{tabularx}

\smallskip
	{\centering
    \begin{tabularx}{0.8\linewidth}{llX}
        Legend: & $n$ & number of vertices in the input graph    \\
        & $k$ & maximum number of waypoints per demand    \\
        & $d$ & number of demands    \\
        & $\tau$ & vertex cover number of the input graph, i.\,e.\ minimum number of vertices to remove to get an edgeless graph    \\
    \end{tabularx}
    }
     \label{tab:summary}
\end{table}

\paragraph*{Our Contributions}

Given its practical importance, our key objective is to search for efficient algorithms solving \sr{} that yield provably optimal results. 
Such algorithms can then be applied directly or used to analyze the quality of faster heuristic approaches.
A folklore brute-force algorithm running in~$O(n^{dk}mkd)$ time is too slow for practical applications:
The number of demands~$d$ is usually quite large and thus should not appear in the exponent. 
In contrast, because of hardware limitations, the number~$k$ of allowed waypoints should be small in practice~\cite{guedrez2017}.
Hence, algorithms running in time~$d^{O(1)}n^k$ would be acceptable, as the dependency on~$d$ is only polynomial.
However, our results exclude (under standard assumptions from computational complexity) algorithms running in time~$f(d) n^{g(k)}$ for any functions~$f$ and~$g$, even exponential or superexponential functions.
More precisely, we show that \sr{} is W[1]-hard with respect to the parameter~$d$ even if~$k = 1$; the above statement is a consequence thereof. 

Another approach to tackle such challenging problems is to leverage the structure of the input graph. 
Indeed, it is reasonable to assume that the well-known graph parameter \emph{treewidth}\footnote{Intuitively, the treewidth of a graph measures how far the graph is from being a tree. A tree has treewidth 1 while a complete graph has treewidth equals to the number of vertices minus 1, the highest possible value.} is reasonably small~\cite{rost2019} (\emph{e.\,g.}, the SNDlib~\cite{sndlib} database provides realistic telecommunication networks, most of which have a treewidth of at most 5). 
Moreover, there is a long list of problems that can be solved efficiently on graphs with small treewidth~\cite{CFKLMPPS15}, including \textsc{Waypoint Routing} \cite{SS22}. 
Unfortunately, we discover that \sr{} is presumably not on this list. We show that \sr{} remains NP-complete on complete bipartite graphs with only four vertices on one side; such graphs have treewidth three.
Hence, a polynomial-time algorithm on bounded-treewidth graphs would imply P${}={}$NP.
Surprisingly, \sr{} turns out to be W[1]-hard even with respect to the number of vertices. 
Hence, algorithms running in time~$2^n \cdot d^{O(1)}$ are presumably impossible as well.
We refer to \cref{tab:summary} for the full picture on the computational complexity of \sr{}. 

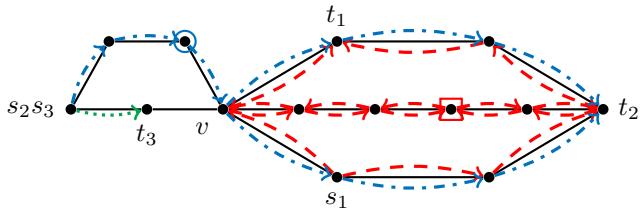
\begin{figure}[t]
    \centering
    \begin{tikzpicture}[yscale=0.9]
        \input{figures/tricky-example}
    \end{tikzpicture}
    \caption{
		A tricky instance for \sr{} where edge weights, edge capacities and bandwidth requirements are one. One waypoint per demand is allowed.
		The demand~$(s_3,t_3)$ blocks an edge and forces a waypoint for demand~$(s_2,t_2)$ between $s_2$ and $v$. The route of $(s_2, t_2)$ then splits between $v$ and $t_2$.
		To ensure that capacities are not exceeded on the lower and upper paths from $v$ to $t_2$, the route of demand~$(s_1, t_1)$ needs to be split too.
		The solution is to select a waypoint for $(s_1, t_1)$ on the middle path from $v$ to $t_2$.
		This means the middle path is used in \emph{both} directions by the route of demand~$(s_1, t_1)$.
    }
    \label{fig:strange-solution}
\end{figure}

There are two sources for these strong intractability results: 
First, one part of \sr{} is related to number packing problems: The demands (numbers) need to be packed (on edges) without exceeding given capacities. 
Indeed, we encode \binpacking{} in \sr{} for the W[1]-hardness with respect to the number of vertices.
There are many approximation algorithms for \binpacking{}~\cite{CCGMV13}, some of which could probably be used as subroutines when dealing with these number issues in \sr{}.
The second source of intractability comes from routing problems: Can we find disjoint paths between given pairs of terminals? 
The approximability of these problems is much less understood~\cite{Kaw11}.
To make progress with the routing issues, we consider \sr{} with unit demands and capacities, which already allows for some counter-intuitive solutions (see \cref{fig:strange-solution}).
This case is trivially solvable on trees and we extend it by providing a polynomial-time algorithm for \sr{} with unit demands and capacities on cacti, a class of graphs with cycles that still follow a clear tree structure: a cactus is a connected graph in which any two simple cycles have at most one vertex in common. In particular, \textsc{Waypoint Routing} is efficiently solvable on them \cite{AFJPS18}. This is an interesting special case for telecommunication networks, as more than one third of the networks in the Internet Topology Zoo dataset are cactus graphs \cite{AFJPS18}.
 
\paragraph*{Paper organization}
The remainder of the paper is organized as follows. 
In Section~\ref{sec:prelim}, we give some notations, formal definitions, and basic observations.   
Section~\ref{sec:lower_bounds_standardparam} presents several NP-hardness and W[1]-hardness results for particular cases where $d$ or $k$ are constant.
Section~\ref{sec:lower_bounds_structuralparam} presents several NP-hardness results for input graphs with constant vertex cover number.
In Section~\ref{sec:cactus_graphs} we study the complexity for the cactus graphs: the problem is NP-hard in general and polynomial-time solvable in the case with unit capacities and demands. Some conclusions are given at the end of the paper.  

\section{Preliminaries \& Basics}
\label{sec:prelim}

\paragraph*{Parameterized Complexity}
We briefly recall the relevant notions of parameterized complexity (see Cygan et al.~\cite{CFKLMPPS15} for a more rigorous introduction). 
A problem is \emph{fixed-parameter tractable} (FPT) with respect to a parameter~$k$ if there is a computable function~$f$ such that any instance~$(\mathcal{I},k)$ can be solved in $f(k)\cdot |\mathcal{I}|^{O(1)}$ time.  
The problem class W[1] is a basic class of presumed parameterized \emph{intractability}. 
A \emph{parameterized reduction} maps an instance $(\mathcal{I},k)$ in $f(k)\cdot |\mathcal{I}|^{O(1)}$ time to an equivalent instance $(\mathcal{I}',k')$ with $k'\le g(k)$ for some computable functions~$f$ and~$g$. 
A parameterized reduction from a W[1]-hard problem~$L$ to a problem~$L'$ proves W[1]-hardness of $L'$ and thus makes FPT algorithms for~$L'$ unlikely.

\paragraph*{Notation}
The notions and notations defined below for our model are inspired by previous works~\cite{aubry2020models, hartert2015solving}.

A \emph{network} is a tuple $(G=(V,E), \omega, c)$ where $G$ is a graph with vertex set $V$ and edge/arc set~$E$, $\omega: E \rightarrow \mathbb{N}$ is a weight function used to compute the shortest paths in $G$, and $c\colon E \rightarrow \mathbb{N}$ is a capacity function.

Depending on the model, the graph can be undirected, bidirected, or directed. When it is bidirected, we assume that an arc has the same weight and capacity as its symmetric.

Let $(G=(V,E), \omega, c)$ be a network. The \emph{forwarding graph} from a \emph{source} $u \in V$ to a \emph{target}~$v \in V$ is the subgraph of $G$ containing all edges/arcs that belong to any shortest path from $u$ to $v$. It is denoted $FG(u,v)$. Even if $G$ is undirected, the forwarding graph $FG(u,v)$ is directed, rooted in~$u$, with edges oriented like the flow from $u$ to $v$.

When $FG(u,v)$ is not a simple path, the \emph{Equal-Cost Multi-Path} (ECMP) mechanism is activated: the incoming flow in a vertex of $FG(u,v)$ is evenly divided between the outgoing edges/arcs of this vertex in $FG(u,v)$. When a forwarding graph is a simple path, we say that it is \emph{ECMP-free}.

A \emph{segment path} in a network $(G=(V,E), \omega, c)$ is a succession of forwarding graphs such that the target of the previous forwarding graph coincides with the source of the next one. A segment path is denoted by $\lsr s, w_1, \ldots, w_\ell, t \rsr$ where $s$ is the source of the segment path, $t$ the target, and~$w_1, \ldots, w_\ell \in V$ are the waypoints in that order. The source and target do not count as waypoints. If every forwarding graph composing it is ECMP-free, we say that the segment path is \emph{ECMP-free}.

The fraction of flow traversing each edge/arc in a forwarding graph that is not ECMP-free can be computed efficiently~\cite{aubry2020models}. See \cref{fig:definitions} for an illustration of ECMP and of a segment path with a waypoint.

A \emph{demand} on a network~$(G=(V,E), \omega, c)$ is a triple~$(s,t,b)$ where the \emph{terminals} $s,t \in V$ are respectively the \emph{source} and \emph{target} of the demand, and $b \in \mathbb{N}$ is the bandwidth requirement, \emph{i.e.} the amount of flow that will be sent from $s$ to $t$. When the bandwidth requirement is 1, we say that the demand is \emph{unit} and we only give the couple $(s, t)$.

Given a network $(G=(V,E), \omega, c)$ and a set of $d$ demands $D = \lbrace (s_i, t_i, b_i) : i=1, \ldots, d \rbrace$ on~$G$, a \emph{routing scheme} for $D$ is a set of $d$ segment paths $\lbrace p_i : i=1, \ldots, d \rbrace$ such that $p_i$ is a segment path from $s_i$ to $t_i$. A routing scheme is \emph{feasible} if the total amount of flow traversing each edge/arc~$e$ of~$G$ is not greater than its capacity $c(e)$.

\paragraph*{Problem Definition} We study \sr{}, which is defined as follows:

\pbdef{\sr}{A network~$(G=(V,E), \omega, c)$, a set of~$d$ demands~$D = \lbrace (s_i, t_i, b_i) : i=1, \ldots, d \rbrace$ and an integer~$k$ called \emph{budget}.}{Is there a feasible routing scheme in~$G$ using at most~$k$ waypoints for each demand?}

We also consider a restriction of the problem called \unitsr where all weights, capacities and bandwidth requirements are equal to 1. \unarysr is the version of \sr{} where the weights, capacities and bandwidth requirements are given in unary encoding.

\begin{observation}
    \sr, \unitsr and \unarysr are in NP on undirected, bidirected, and directed graphs.
\end{observation}
\begin{proof}
    Consider a certificate for \sr{} consisting of a set of segment paths, one per demand, each encoded by $\lsr s_i, w_1, \ldots, w_\ell, t_i \rsr$. We can check in time $O(kd)$ that each of them uses at most $k$ waypoints. What is left is to check if some edge/arc capacity is exceeded, which can be done by calling a shortest path algorithm a polynomial number of times. It follows that \unitsr and \unarysr are also in NP.
\end{proof}

\begin{lemma} [folklore]
\label{lem:folklore}
    \sr, \unitsr and \unarysr on undirected, bidirected, and directed graphs can be solved in time $O(n^{kd}mkd)$, where $n$ is the number of vertices and $m$ is the number of edges/arcs.
\end{lemma}
\begin{proof}
    We can assume that the forwarding graph and associated flow function have been efficiently computed for every pair of vertices \cite{aubry2020models}. 
    Then the brute force algorithm works as follows: 
    There are $n^{kd}$ routing schemes. For each of them, one can check in $O(mkd)$ that the induced load on every edge/arc does not exceed its capacity: there are $m$ arcs/edges and at most $(k+1)d$ forwarding graphs.
\end{proof}

\section{Parameters in the Input}
\label{sec:lower_bounds_standardparam}

We define some gadgets used in the two next reductions, where we consider cases with a high budget~$k$ of waypoints. 

\begin{definition}
    Let $G=(V,E)$ be an undirected or directed graph. The $\ell$-\emph{extended graph} of $G$, denoted $S_\ell(G)$, is obtained by replacing every edge/arc $e \in E$ by a path of $\ell$ edges/arcs.
\end{definition}

If two vertices $u,v \in V$ are at distance $d_G(u,v)$ in $G$ in number of edges/arcs, then they are at distance $\ell d_G(u,v)$ in $S_\ell(G)$. In particular, the distance between any two original vertices of $G$ is at least $\ell$ in $S_\ell(G)$.

\begin{figure}[t]
    \centering
    \begin{tikzpicture}[scale=0.8]
    \input{figures/triangle-chain}
    \end{tikzpicture}
    \caption{The gadget $\nabla_4$ in a network with unit edge weights and capacities, with two demands $(s_1,t_1)$ and $(s_2,t_2)$ traversing it. Square waypoints are for demand 1 and round waypoints are for demand 2.}
    \label{fig:triangle_chain}
\end{figure}
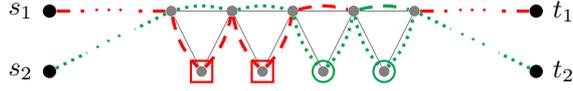

\begin{definition}
    A \emph{triangle chain} of length $\ell$, denoted $\nabla_\ell$, is an undirected graph made up of $\ell$ cliques of size 3 (triangles) attached to each other by shared vertices, as shown in \cref{fig:triangle_chain}.
\end{definition}

In a network where all edge weights and capacities are 1, routing two demands through chain $\nabla_{2\ell}$ requires a budget $k \geq \ell$, and when $k=\ell$, each demand has exactly $k$ waypoints in $\nabla_{2 k}$: in each triangle, one demand can use the top edge (shortest path) and the other must be diverted by a waypoint placed in the bottom vertex of the triangle, as shown in \cref{fig:triangle_chain}.

\begin{theorem}
\label{thm:twoedp}
    \unitsr is NP-complete on directed graphs even for $d=3$.
\end{theorem}

\begin{proof}
    We provide a polynomial-time reduction from the NP-complete problem~\twoedplong (\twoedp) on directed graphs~\cite{fortune1980directed}.

    \pbdef{\twoedplong (\twoedp) on directed graphs}{A directed graph $G=(V,E)$ and two pairs of vertices $(s_1, t_1), (s_2, t_2)$.}{Are there two arc-disjoint paths $P_1, P_2$ from $s_1$ to $t_1$ and from $s_2$ to $t_2$ respectively?}
    
    Let $\mathcal{I}=(G=(V,E), D)$ be an instance of \twoedp on a directed graph. We construct in polynomial time an instance $\mathcal{I'}=(G'=(V', E'), D', k)$ of \unitsr on a directed graph as follows (see \cref{fig:red_2EDP}):
	\begin{figure}[t]
		\centering
		\begin{tikzpicture}[scale=0.6]
        \input{figures/red-from-2edp}
		\end{tikzpicture}
		\caption{Illustration of the reduction from \twoedp to \unitsr: on the left is the reduced instance and on the right its demand graph.}
		\label{fig:red_2EDP}
	\end{figure}
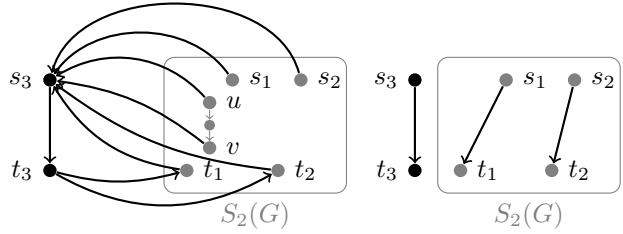
    \begin{itemize}
        \item Start with the 2-extended graph $S_2(G)=(V',E')$.
        \item Add two new vertices $s_3, t_3$ in $V'$.
        \item For each original vertex $v \in V$ in $S_2(G)$, add the arc $v s_3$ in $E'$. Add the arcs $t_3 t_1$, $t_3 t_2$ and $s_3 t_3$.
        \item Set $D'= \lbrace (s_1, t_1), (s_2, t_2), (s_3, t_3) \rbrace$.
        \item Set $k=|V|$.
    \end{itemize}
    Clearly the above transformation is polynomial. We show that $\mathcal{I}$ is a yes-instance if and only if $\mathcal{I'}$ is a yes-instance.

    ``$\Rightarrow$'' Let $P_1, P_2$ be a solution to $\mathcal{I}$. We assume they are simple paths. Denote $P_1$ by its arcs in $G$: $P_1=(a_1, \ldots, a_{\ell_1})$. Denote $v_i$ the vertex in the middle of the path of two arcs that replaced $a_i$ in $S_2(G)$. Then demand $(s_1, t_1)$ uses $\lsr s_1, v_1, \ldots, v_{\ell_1}, t \rsr$. Proceed identically for $(s_2, t_2)$. Demand $(s_3,t_3)$ uses $\lsr s_3, t_3 \rsr$ (the single arc $s_3 t_3$).

    ``$\Leftarrow$'' Conversely, suppose we have a feasible routing scheme for $\mathcal{I'}$. 
    Since $s_3 t_3$ is the only arc leaving $s_3$, the only way to route $(s_3, t_3)$ is to use $\lsr s_3, t_3 \rsr$ and it saturates the arc. It implies that in any feasible routing scheme, the segment paths of $(s_1, t_1)$ and $(s_2, t_2)$ are contained in $S_2(G)$. 
    Then we prove that in any feasible routing scheme, these two segment paths are ECMP-free. Targeting a contradiction, let $u,v$ be two vertices of $S_2(G)$ such that $FG(u,v)$ is contained in $S_2(G)$ and is not ECMP-free. We can assume that $u,v \in V$ because the other vertices of $S_2(G)$ have in- and out-degree one. There are no parallel arcs in $G$, so two distinct shortest paths from $u$ to $v$ should have length at least 2 in $G$ and 4 in $S_2(G)$. However, the path from $u$ to $v$ using the shortcut $s_3 t_3$ in $G'$ has length 3, so we have a contradiction. 
    Thus, the segment paths of $(s_1, t_1)$ and $(s_2,t_2)$ are ECMP-free. They are also arc-disjoint since capacities are respected, so we can translate them into arc-disjoint paths in $G$. 
\end{proof}

\begin{theorem}
\label{thm:twodonep}
    \unitsr is NP-complete on undirected graphs even for $d=4$.
\end{theorem}

\begin{proof}
    We provide a polynomial-time reduction from the NP-complete problem \twodoneplong (\twodonep) on undirected graphs~\cite{eilam1998disjoint}. 

    \pbdef{\twodoneplong (\twodonep)}{An undirected graph $G$, two pairs of vertices $(s_1, t_1)$ and $(s_2, t_2)$.}{Are there two edge-disjoint paths $P_1$ from $s_1$ to $t_1$ and $P_2$ from $s_2$ to $t_2$ such that $P_1$ is a shortest path?}

    Let $\mathcal{I} = (G=(V,E), s_1, t_1, s_2, t_2)$ be an instance of \twodonep, $G$ connected. Denote $n=|V|$. We construct in polynomial time an instance $\mathcal{I'}=(G', D, k)$ of \unitsr as follows (see \cref{fig:red_2D1SP}):
	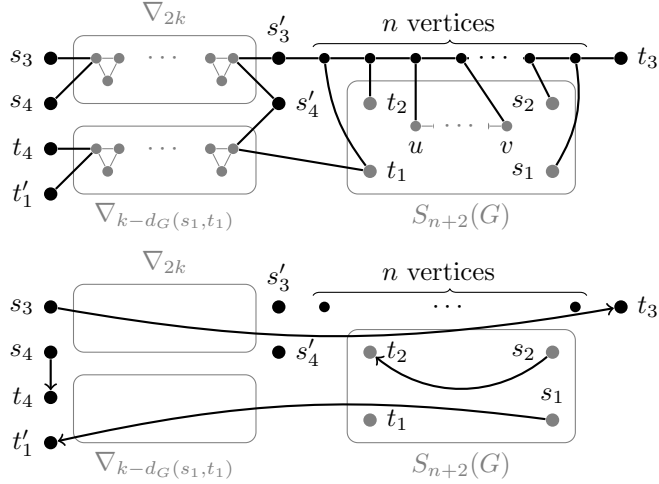
\begin{figure}[t]
		\centering
		\begin{tikzpicture}[scale=0.6]
        \input{figures/red-from-2d1sp1}
		\end{tikzpicture}
        \vspace{1em}
		\begin{tikzpicture}[scale=0.6]
        \input{figures/red-from-2d1sp2}
		\end{tikzpicture}
		\caption{Illustration of the reduction from \twodonep to \unitsr. The reduced instance is at the top and below is its demand graph.}
		\label{fig:red_2D1SP}
	\end{figure}
    \begin{itemize}
        \item Start with the $n+2$-extended graph $S_{n+2}(G)=(V',E')$. 
        \item Set the budget to $k=n$.
        \item Add a triangle chain $\nabla_{2k}$. Connect one end of this chain to two new vertices $s_3$ and $s_4$ and the other end to two new vertices $s_3'$ and $s_4'$.
        \item Create a \emph{shortcut} path of $n$ vertices and connect each of them to a different original vertex of $V$ in $S_{n+2}(G)$. Connect one end of the shortcut path to $s_3'$ and the other end to a new vertex $t_3$.
        \item Add a triangle chain $\nabla_{k-d_G(s_1,t_1)}$ where $d_G(s_1,t_1) \leq n$ is the distance from $s_1$ to $t_1$ in $G$ (value computable in polynomial time). Connect one end of this chain to $t_1$ and $s_4'$ and the other end to two new vertices $t_1'$ and $t_4$.
        \item Set $D=\lbrace (s_1, t_1'), (s_2, t_2), (s_3, t_3), (s_4, t_4) \rbrace$.
    \end{itemize}
    Clearly the above transformation is polynomial. We show that $\mathcal{I}$ is a yes-instance if and only if $\mathcal{I'}$ is a yes-instance.

    ``$\Rightarrow$'' Let $P_1$, $P_2$ be a solution of $\mathcal{I}$. We assume they are simple paths. 
    
    First assign $k$ waypoints to $(s_3, t_3)$ and $k$ waypoints to $(s_4, t_4)$ to route them through $\nabla_{2k}$ as described in \cref{fig:triangle_chain}. Having used all their budget in $\nabla_{2k}$, when leaving $\nabla_{2k}$, $(s_3, t_3)$ uses the straight line across the shortcut path to reach $t_3$, and $(s_4, t_4)$ uses the straight line across $\nabla_{k-d_G(s_1, t_1)}$ to reach $t_4$.
    
    Denote $P_2$ by its edges: $P_2=(e_1, \ldots, e_\ell)$. Denote $v_i$ the middle vertex of the path of length $n+2$ that replaced $e_i$ in $S_{n+2}(G)$ (if $n+2$ is odd, take $v_i$ as the middle vertex that is closest to the origin of the path). Then route demand $(s_2, t_2)$ on $\lsr s_2, v_1, \ldots, v_\ell, t_2 \rsr$, using $\ell \leq k$ waypoints. This segment path stays in $S_{n+2}(G)$ because the shortest path between two consecutive endpoint/waypoints is the one not using the shortcut. The segment path of $(s_1, t_1')$ is defined similarly between $s_1$ and $t_1$ using $P_1$ ($t_1$ is not a waypoint), using exactly $d_G(s_1, t_1)$ waypoints in $S_{n+2}(G)$. The rest of the segment path is given by $k-d_G(s_1, t_1)$ waypoints in $\nabla_{k-d_G(s_1, t_1)}$. The shortest path between the last waypoint in $S_{n+2}(G)$ and the first in $\nabla_{k-d_G(s_1, t_1)}$ does not use the shortcut and traverses $t_1$ because it is the only way to reach $t_1'$.

    ``$\Leftarrow$'' Conversely, suppose we have a feasible routing scheme for $\mathcal{I'}$. Demands $(s_3, t_3)$ and $(s_4, t_4)$ require $k$ waypoints each in $\nabla_{2k}$, so after leaving $\nabla_{2k}$ they saturate respectively the shortcut path from $s_3'$ to $t_3$ and the straight line from $s_4'$ to $t_4$ across $\nabla_{k-d_G(s_1,t_1)}$. It implies that the segment paths of $(s_1, t_1')$ (from $s_1$ to $t_1$) and $(s_2,t_2)$ are contained in $S_{n+2}(G)$, and that the segment path of $(s_1, t_1')$ uses at most $d_G(s_1, t_1)$ waypoints between $s_1$ and $t_1$.
    
    Notice that a path in $S_{n+2}(G)$ between two vertices $s,t \in V$ is made of subpaths of length $n+2$ corresponding each to an edge of $G$. Call these subpaths \emph{extended edges}. Let $u,v$ be two vertices of $S_{n+2}(G)$. If $u,v \in V$, then a shortest path from $u$ to $v$ uses the shortcut, with length at most $n+1$. If $u \in V$ and $v \notin V$, then unless $v$ belongs to an extended edge incident to $u$, a shortest path between them also uses the shortcut. Finally, if $u, v \notin V$, then unless $u$ and $v$ are located on adjacent extended edges or the same extended edge, a shortest path between them uses the shortcut.

    Consider the segment path of $(s_2, t_2)$ in our feasible routing scheme: $\lsr s_2, w_1, \ldots, w_\ell, t_2 \rsr$. Given the remarks above, we deduce that there is at least one waypoint per extended edge. It implies that this segment path is ECMP-free, and it can be translated into a solution path for $\mathcal{I}$. Consider the segment path of $(s_1, t_1')$ in our feasible routing scheme. It needs at least $k-d_G(s_1, t_1)$ waypoints in $\nabla_{k-d_G(s_1, t_1)}$, leaving at most $d_G(s_1, t_1)$ waypoints for the part in $S_{n+2}(G)$. Given the above, we also deduce that it uses at least $d_G(s_1, t_1)$ waypoints between $s_1$ and $t_1$ and that it is ECMP-free too. Since a path from $s_1$ to $t_1$ uses at least $d_G(s_1, t_1)$, and that it is possible to use exactly one waypoint per extended edge as shown in $\Rightarrow$, this section of the segment path can be translated into a path of length exactly $d_G(s_1, t_1)$ in $G$, so we have a shortest path for $\mathcal{I}$.
\end{proof}

We now consider the case where~$k=1$ and~$d$ is small (but not constant).

\begin{theorem}
	\label{thm:mcc}
    \unitsr parameterized by the number of demands is W[1]-hard on undirected, bidirected, and directed graphs even for $k=1$.
\end{theorem}

We provide a parameterized reduction from the W[1]-complete problem \mcc parameterized by the number of colors~\cite{pietrzak2003parameterized}. 
We significantly modify a reduction by \citet{bentert2023using} to obtain our result. 

\pbdef{\mcc}{An undirected graph~$G=(V,E)$, an integer~$d$, a coloring function~$c\colon V \rightarrow \lbrace 1, \ldots, d \rbrace$.}{Is there a multicolored clique in $G$, that is, a set of pairwise adjacent vertices containing exactly one vertex of each color?}

Let $\mathcal{I}=(G=(V, E), d, c)$ be an instance of \mcc parameterized by $d$. For each $i=1, \ldots, d$ denote $V_i = \lbrace v \in V : c(v)=i \rbrace$. Then $(V_1, \ldots, V_d)$ is a partition of $V$. We make the following assumptions:
\begin{enumerate}
	\item The $d$ sets $V_1, \ldots, V_d$ have equal size $n \geq 2$: otherwise, take $n=\max_i |V_i|$ and add isolated vertices in the sets of size $<n$. The case $|V_i|=1 \quad \forall i$ is trivial so we assume $n \geq 2$.
	\item The number of colors $d$ is even: if it is odd, add a new color, create $n$ vertices of that color, connect one of these new vertices to every $v \in V_1 \cup \ldots \cup V_d$ and leave the other $n-1$ new vertices isolated.
\end{enumerate}

\paragraph{Construction of the reduced instance}

\begin{figure}[t]
		\centering
		\begin{tikzpicture}[scale=0.8]
        \input{figures/red-from-mcc-layout}
		\end{tikzpicture}
		\caption{
			General layout of the reduced instance: 
			The horizontal and vertical paths are grouped by the colors of the vertices the paths represent.
			$L$-separators are represented by dashed lines and $\ell$-subpaths by dotted lines (shortcuts are not represented).
		}
		\label{fig:red_mcc1}
	\end{figure}
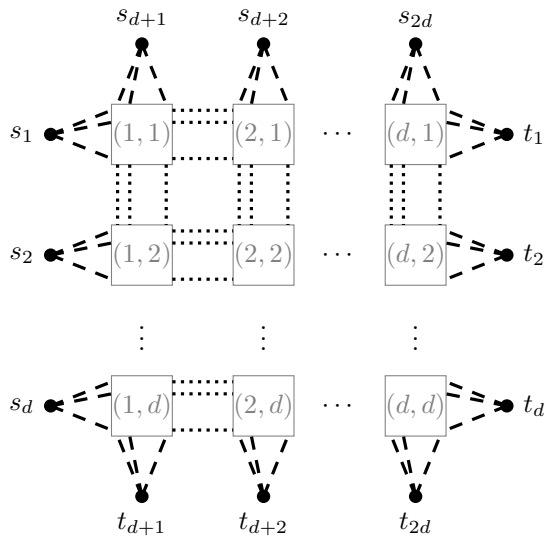

\begin{figure}[t]
	\centering
	\begin{tikzpicture}[scale=0.9]
    \input{figures/red-from-mcc-example1}
	\end{tikzpicture}
	\begin{tikzpicture}[scale=0.35]
	\input{figures/red-from-mcc-example2}
	\end{tikzpicture}
	\caption{Illustration of the reduction from \mcc to \unitsr: a dummy color has been added in the instance of \mcc and not all shortcuts are represented in the reduced instance. Dashed lines represent $L$-separators and dotted lines represent $\ell$-subpaths. The clique in the figure on the left corresponds to the highlighted paths in the reduced instance.}
	\label{fig:red_mcc3}
\end{figure}
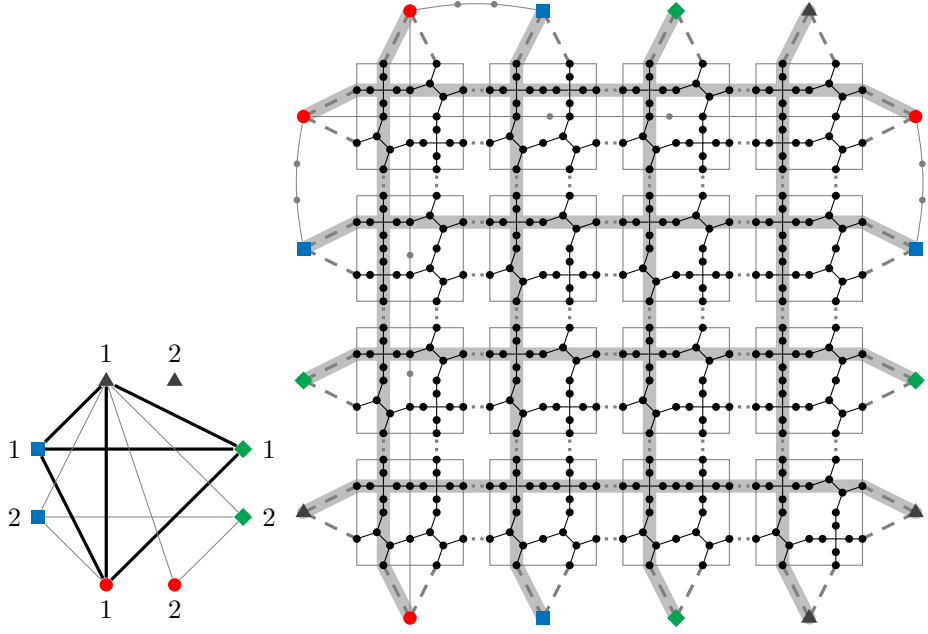

    We construct in polynomial time an instance $\mathcal{I}'=(G'=(V',E'), D, k)$ of \unitsr on an undirected graph parameterized by the number of demands as follows (see \cref{fig:red_mcc3}):
	Start with an empty graph $G'$. For each color $i= 1,\ldots, d$, add four vertices in $V'$: $s_i$ and $t_i$ facing each other, $s_{d+i}$ and $t_{d+i}$ facing each other.
	For each vertex $v \in V$, create a horizontal path $P_v$ from $s_{c(v)}$ to $t_{c(v)}$ and a vertical path $Q_v$ from $s_{d+c(v)}$ to $t_{d+c(v)}$. 
	The paths are arranged in a grid and in the same order of $V$ horizontally and vertically, grouped by color; see \cref{fig:red_mcc1} for an illustration.
	
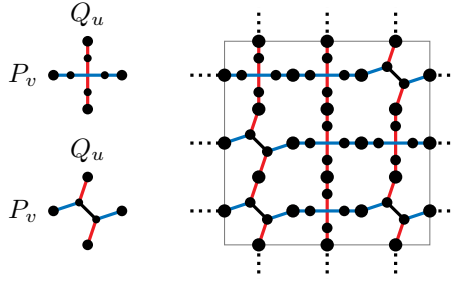
\begin{figure}[t]
	\centering
	\begin{tikzpicture}[scale=0.45]
    \input{figures/red-from-mcc-intersections}
	\end{tikzpicture}
	\caption{The square shows what is represented by squares in \cref{fig:red_mcc1}.
        \emph{Left side:} Two ways that paths~$P_v$ and~$Q_u$ can intersect. 
		The top refers to when~$u$ and~$v$ can be in a clique together: the subpaths of~$P_v$ and~$Q_u$ do not share any edge. Thus, they can be used in a routing scheme together.
		Below is the case where~$u$ and~$v$ cannot be in a clique together: the two subpaths intersect and it is not possible to use both of them in the routing scheme.
		\emph{Right side:} An example for $n=3$ of a complete color overlap between two colors.}
	\label{fig:red_mcc2}
\end{figure}

	A path $P_v$ (resp. $Q_v$) consists of several subpaths as follows:
	\begin{itemize}
		\item in the middle, alternately $d$ subpaths of $3n$ edges and $d-1$ subpaths of $\ell=16$ edges, called \emph{$\ell$-subpaths},
		\item between $s_{c(v)}$ (resp. $s_{d+c(v)}$) and the first subpath of length $3n$, a subpath of $L=2\big(3nd+(d-1)\ell \big)$ edges called \emph{$L$-separator}, and another one between the last subpath of length $3n$ and $t_{c(v)}$ (resp. $t_{d+c(v)}$).
	\end{itemize}
	The length of a $L$-separator is chosen so that it is twice as long as the subpath between the two $L$-separators.
	The first subpath of length $3n$ corresponds to where $P_v$ overlaps with the $n$ vertical paths corresponding to~$V_1$, the second corresponds to where $P_v$ overlaps with vertical paths corresponding to~$V_2$, and so on.
	We call these parts \emph{color overlap}; these refer to the boxes in \cref{fig:red_mcc1}.
	More precisely, given two colors $i, j$, the \emph{color overlap} $(i,j)$ refers to the area where the paths $P_v, v \in V_i$ and $Q_u, u \in V_j$ overlap. 
	Merge some edges in the color overlaps: for all $u,v \in V$, $u \neq v$, the crossover of $Q_u$ and $P_v$ consists of three horizontal edges and three vertical edges. 
	If $c(u) = c(v)$ or $uv \notin E$, then $u$ and $v$ cannot be together in a clique, so merge the middle edge as shown in \cref{fig:red_mcc2}.
	
	Add a path of three edges, called \emph{shortcut}, between each of the following pairs of endpoints:
	\begin{itemize}
		\item $\lbrace s_i, t_i \rbrace$ and $\lbrace s_{d+i}, t_{d+i} \rbrace$ for $1 \leq i \leq d$ (one shortcut per pair of terminals facing each other),
		\item $\lbrace s_i, s_j \rbrace$, $\lbrace t_i, t_j\rbrace$, $\lbrace s_{d+i}, s_{d+j} \rbrace$, $\lbrace t_{d+i}, t_{d+j} \rbrace$ for $1 \leq i < j \leq d$ (one shortcut per pair of terminals located on the same side of the grid).
	\end{itemize}
	Given a pair $\lbrace u,v \rbrace$ above, we denote by~$u$-$v=v$-$u$ the associated shortcut.
	For each $i=1,\ldots,d$, create demands $(s_i, t_i)$ and $(s_{d+i}, t_{d+i})$ in $D$. For each shortcut, create a \emph{blocker demand} in $D$ corresponding to the middle edge of the shortcut path.
	Finally, set $k=1$.

The above transformation is polynomial.
The parameter of the reduced instance $\mathcal{I}'$ is the number of demands $d'=|D|=2d^2+2d$.

\paragraph{Properties of the reduced instance}
Note that since $k=1$, we are only interested in shortest paths where at least one endpoint is a terminal. 
The intuition about how shortest paths behave in the reduced instance is as follows:
\begin{itemize}
	\item The length $L$ is twice as large as the grid, so when possible, a shortest path does not contain any $L$-separator. 
	Any shortest path from a vertex within the grid to a terminal is of length $<2L$ so it contains exactly one $L$-separator. 
	Any shortest path between two terminals that are on adjacent sides of the grid contains exactly two $L$-separators.
	\item The $\ell$-subpaths separate color overlaps from each other inside the grid. The length $\ell$ is longer than using a shortcut between terminals of the same side outside of the grid.
	\item In a color overlap in which some edges are merged, it is possible to go from a line to another (up or down), however it costs at least one more edge compared to using a straight line. The same goes for columns. Thus, a shortest path will avoid changing lines or columns inside the grid and rather go straight to or from a terminal.
\end{itemize}
Overall, a shortest path between two vertices will be one that minimizes first the number of $L$-separators, then the number of $\ell$-subpaths, and finally the number of turns inside the grid.

\begin{lemma}
\label{lem:mcc1}
	It is possible to represent any path $P_v$ or $Q_v$ as a segment path with a single waypoint placed in one of the three middle vertices of $P_v$ or $Q_v$.
\end{lemma}
\begin{proof}
	Note that for all $v \in V$, $P_v$ and~$Q_v$ have even length~$2L+3nd+(d-1) \ell$ as~$d$ and~$\ell$ are even. 
	Denote by~$p_{v}$ (respectively~$q_{v}$) the vertex located exactly in the middle of~$P_{v}$ (respectively~$Q_{v}$). They are both exactly in the middle of an $\ell$-subpath. The segment path $\lsr s_{c(v)}, p_v, t_{c(v)} \rsr$ (resp. $\lsr s_{d+c(v)}, q_v, t_{d+c(v)} \rsr$ is exactly $P_v$ (resp. $Q_v$): the unique shortest path from $s_{c(v)}$ to $p_{v}$ is the first half of $P_{v}$ and the unique shortest path from $p_{v}$ to $t_{c(v)}$ is the second half of $P_{v}$. The same reasoning applies to $\lsr s_{d+c(v)}, q_{v}, t_{d+c(v)} \rsr$ and $Q_{v}$.
	It also works with the two vertices on each side of the middle vertex as waypoint because the shortcut $s_{c(v)}$-$t_{c(v)}$ has length 3.
\end{proof}

\begin{lemma}
	\label{lem:mcc2}
	A shortest path from a terminal $s_i \in \lbrace s_1, \ldots, s_d\rbrace$ to a vertex $u$ belonging to some path $P_v$ with $v \notin V_i$ contains a shortcut.
\end{lemma}
\begin{proof}
	Denote $j=c(v)$. A shortest path from $s_i$ to $u$ uses at least one shortcut among $s_i$-$t_i$, $s_i$-$s_j$, $s_i$-$t_i$ and $t_i$-$t_j$, because it avoids traversing additional $\ell$-subpaths.
\end{proof}

\begin{lemma}
	\label{lem:mcc3}
	A shortest path between two terminals located on adjacent sides of the grid contains a shortcut unless they are located in the same corner of the grid.
\end{lemma}
\begin{proof}
	Let $s_i$ be a terminal on the left side and $w$ be a terminal on the top or bottom side. The case with a terminal on the right side is similar. Any shortest path from $s_i$ to $w$ contains exactly two $L$-separators and has to visit at least one color overlap. Notice that there exists a path of length $<2L+\ell$ from $s_i$ to any terminal on the top or bottom side, traversing color overlap $(1,1)$ and using up to three shortcuts. So any shortest path from $s$ to $w$ does not contain a $\ell$-subpath and thus visits exactly one corner color overlap $(1,1), (1,d), (d,1)$ or $(d,d)$. Unless $(s_i, w)=(s_1, s_{d+1})$ or $(s_i, w)=(s_d, t_{d+1})$, it contains a shortcut.
\end{proof}

\paragraph{Solution equivalence}
    
\begin{lemma}
	\label{lem:w[1]-hard-correct}
	The original instance $\mathcal{I}$ is a yes-instance if and only if the reduced instance $\mathcal{I}'$ is a yes-instance.
\end{lemma}

\begin{proof}
    
    ``$\Rightarrow$'': 
    Suppose that there is a multicolored clique~$S=\lbrace v_1, \allowbreak\ldots, v_d \rbrace$ in~$G$ where~$c(v_i)=i$. 
    We construct a solution for $\mathcal{I}'$. 
    Assign no waypoint to the blocker demands; they all use the middle edge of the shortcut they are located on. Using \cref{lem:mcc1}, take the path $P_{v_i}$ for demand~$(s_i, t_i)$ and~$Q_{v_i}$ for~$(s_{d+i}, t_{d+i})$.
    These segment paths do not use any shortcut and since $S$ is a clique, they are pairwise edge-disjoint.
    
    ``$\Leftarrow$'': Conversely, assume we have a feasible routing scheme for~$\mathcal{I}'$. 
    First note that every blocker demand saturates at least one edge of the shortcut where it is located. Thus, we can assume that they have no waypoint and saturate the middle edge of every shortcut. It means that the segment path of any non-blocker demand~$(s_i, t_i)$ or~$(s_{d+i}, t_{d+i})$ is disjoint from any shortcut, and it also implies that they all have exactly one waypoint assigned.
    Consider a demand~$(s_i, t_i)$ and denote~$w_i$ its waypoint. We claim that the segment path $\lsr s_i, w_i, t_i \rsr$ is exactly one of the paths $P_v$, $v \in V_i$. 
    To prove the claim, we show that if $w_i$ is one of the three middle vertices of some $P_v, v \in V_i$, then $\lsr s_i, w_i, t_i \rsr$ is exactly $P_v$, and if $w_i$ is another vertex of $G'$, then $\lsr s_i, w_i, t_i \rsr$ contains at least one shortcut. The first part is proved in \cref{lem:mcc1}.  

     Assume now that $w_i$ is not one of these vertices. A waypoint $w_i$ cannot be on a shortcut because the segment path would either contain a shortcut or overload an edge of the shortcut with demand $(s_i, t_i)$ alone. We differentiate the following cases: 
    \begin{enumerate}
        \item \label{mcc1} $w_i$ belongs to a path $P_v$ with $v \in V_i$ but is not one of the three middle vertices,
        \item \label{mcc2} $w_i$ belongs to a path $P_v$ with $v \notin V_i$,
        \item \label{mcc3} $w_i$ is a terminal on the right or left side,
        \item \label{mcc4} $w_i$ is a terminal on the top or bottom side,
        \item \label{mcc5} $w_i$ belongs to a path $Q_v$ only.
    \end{enumerate}
    
    In case \ref{mcc1}, $w_i$ is either closer to $s_i$ or to $t_i$. If it is closer to $s_i$, then the shortest path from $w_i$ to $t_i$ contains the shortcut $s_i$-$t_i$. If $w_i$ is closer to $t_i$, the shortest path from $s_i$ to $w_i$ contains $s_i$-$t_i$. 

    Case \ref{mcc2} is covered by \cref{lem:mcc2}.

    In case \ref{mcc3}, if $w_i$ is on the left side, then the shortest path from $s_i$ to $w_i$ is $s_i$-$w_i$, and if it is on the right side, the shortest path from $w_i$ to $t_i$ is $w_i$-$t_i$.

    In case \ref{mcc4}, $w_i$ cannot be both in the same corner as $s_i$ and $t_i$, so according to \cref{lem:mcc3}, either the shortest path from $s_i$ to $w_i$ or from $w_i$ to $t_i$ contains a shortcut.
    
    In case \ref{mcc5}, $w_i$ is a non-terminal vertex belonging only to a path $Q_v$ so it has degree 2. Consider the two closest vertices of $Q_v$ of degree at least 3 on each side of $w_i$, denoted $q$ and $q'$. They are either terminals, or belong to some path $P_u$. They cannot both be terminals as $Q_v$ has merged edges at least with paths $P_{v'}: c(v')=c(v)$. 
    Any shortest path to $w_i$ must traverse $q$ or $q'$. 
    Since $d$ is even, $Q_v$ is strictly either in the left half of the graph or in the right side. Assume it is on the left side, closer to $s_i$. Let $q$ be one of these two vertices such that a shortest path from $w_i$ to $t_i$ traverses $q$. The subpath from $q$ to $t_i$ is also a shortest path. If $q$ belongs to a path $P_u$, then we are in case \ref{mcc1} or \ref{mcc2} so a shortest path from $q$ to $t_i$ uses a shortcut. If $q$ is a terminal, it cannot be in a corner with $t_i$ so according to case \ref{mcc4} the shortest path from $q$ to $t_i$ uses a shortcut. The reasoning is the same when~$Q_v$ is strictly on the right side of the grid, by considering the shortest path between~$s_i$ and~$w_i$.

    When considering a demand~$(s_{d+i}, t_{d+i})$ and its segment path~$\lsr s_{d+i}, w_{d+i}, t_{d+i} \rsr$, the reasoning is the same: it correspond to a single path $Q_v$. Since the routing scheme is feasible, the segment paths are pairwise edge-disjoint. The set of vertices associated with the segments paths of the demands $(s_i, t_i), i=1, \ldots, d$ gives a multicolored clique in $\mathcal{I}$. The other set associated with the segment paths of demands~$(s_{d+i}, t_{d+i})$ also gives a solution for~$\mathcal{I}$, possibly different.
\end{proof}

\paragraph{Adaptation to bidirected and directed graphs}
    
The reduction can be adapted to bidirected and directed graphs. For bidirected graphs, the construction is the same, but there are two blocker demands on the shortcut between two terminals on the same side, to saturate the shortcut both ways. The parameter of the reduced instance is~$d'=|D|=4d^2$. For directed graphs, the arcs of the grid and the arcs of the shortcuts between two terminals facing each other are directed from left to right and from top to bottom. The shortcuts between two terminals on the same side have both directions and a blocker demand for each direction, so~$d'=4d^2$ too.

\section{Structural Parameters}
\label{sec:lower_bounds_structuralparam}
In the previous section, we saw that \unitsr{} remains hard if~$k$ or~$d$ is small.
Hence, it is natural to also consider the structure of the infrastructure network.
To this end, we consider the case where the vertex cover number of the input graph is constant, that is, every edge in the graph is incident to one of four vertices.

\begin{theorem}
	\label{thm:threeedgecol}
    \unitsr is NP-complete on undirected graphs even with vertex cover number $\tau=4$ and $k=1$.
\end{theorem}

\begin{proof}
    We provide a polynomial-time reduction from the NP-complete problem \threeedgecol~\cite{stockmeyer1973planar}, inspired by a reduction by \citet{fleszar2018new}.

    \pbdef{\threeedgecol}{An undirected graph $G=(V,E)$.}{Is there a proper edge coloring of $G$, i.\,e.\ an edge coloring such that no two adjacent edges have the same color, using at most 3 colors?}

    Let $\mathcal{I} = (G=(V,E))$ be an instance of \threeedgecol. We assume that $G$ is connected, otherwise each connected component can be treated independently. We construct in polynomial time an instance $\mathcal{I'}=(G'=(V',E'), D, k)$ of \unitsr as follows (see \cref{fig:red_3EC}): 

    \begin{figure}[t]
		\centering
		\begin{tikzpicture}[scale=0.7] 
			\input{figures/red-from-3ec}
		\end{tikzpicture}
		\caption{
			Illustration of the reduction from \threeedgecol to \unitsr. 
			Subfigure A is a \threeedgecol instance on four vertices~$1,\ldots,4$ with a solution, subfigure B is the associated reduced instance of \unitsr and its demand graph, and subfigure C is the routing scheme associated with the solution of A. 
			Solid black arc/edges correspond to dummy demands.
		}
		\label{fig:red_3EC}
    \end{figure}
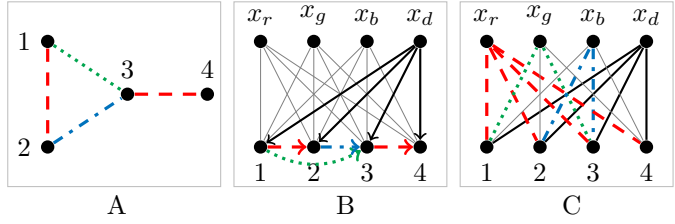
    
    Set $V'= V \cup W$ where $W = \lbrace x_r, x_g, x_b, x_d \rbrace$ contains four additional vertices corresponding to colors ``red'', ``green'', ``blue'', and a ``dummy color''. Make $G'$ a complete bipartite graph between $V$ and $W$.
    For each edge $uv \in E$, create a demand $(u,v)$ in $D$. For each $u \in V$, create a \emph{dummy demand} $(x_d, u)$ in $D$.
    Set $k=1$.
    Clearly, the above transformation is polynomial. 

	We show that $\mathcal{I}$ is a yes-instance if and only if $\mathcal{I'}$ is a yes-instance.
 
    ``$\Rightarrow$'': Let $c\colon E \rightarrow \lbrace r, g, b \rbrace$ be a proper 3 edge coloring of $G$. For each dummy demand $(x_d, v)$, choose the segment path $\lsr x_d,v \rsr$ (no waypoint): it uses only the edge $x_d v$. For every remaining demand $(u, v)$, choose the segment path $\lsr u, x_{c(uv)}, v \rsr$: it uses edges $u x_{c(uv)}$ and $v x_{c(uv)}$. Clearly, a dummy demand does not share an edge with another dummy demand nor with a demand corresponding to an edge in $E$. Consider two demands $(u,v)$ and $(u',v')$ corresponding to edges $uv$ and $u'v'$ in $G$. If the edges are not adjacent, the two segment paths $\lsr u, x_{c(uv)}, v \rsr$ and $\lsr u', x_{c(u'v')}, v' \rsr$ share no edge. If the edges are adjacent, then $x_{c(uv)} \neq x_{c(u'v')}$ so the two segment paths share no edge. Since each edge is used by at most one demand, edge capacities are respected.
    
    ``$\Leftarrow$'': Conversely, suppose that we have a feasible routing scheme for $D$ in $G'$. Notice that there are four shortest paths between two $u,v \in V$, $|V|$ shortest paths between two $x_c, x_{c'} \in W$, and a single shortest path between $v \in V, x_c \in  W$. First we show that in any feasible routing scheme, a dummy demand $(x_d, v)$ must be routed on $\lsr x_d, v \rsr$: assume another segment path $\lsr x_d, w, v \rsr$. If $w \in \lbrace x_r, x_g, x_b \rbrace$, it induces a load of $1 + 1/|V|$ on $w v$. If $w \in V \backslash v$, it induces a load of $5/4$ on $x_d w$. Then we show that in any feasible routing scheme, every demand $(u,v) \in V^2 \cap D$ must have exactly one of $\lbrace x_r, x_g, x_b \rbrace$ as waypoint: If no waypoint is used or if $x_d$ is the waypoint, then $x_d u$ and $x_d v$ are overloaded since dummy demands $(x_d, u)$ and $(x_d, v)$ already use them. If some $w \in V \backslash \lbrace u,v \rbrace$ is the waypoint, then $x_d u, x_d v, x_d w$ are overloaded again because of dummy demands.
    
    Consider the coloring of $E$ associated with the feasible routing scheme. The waypoint of a non-dummy demand is in $\lbrace x_r, x_g, x_b \rbrace$, so it uses at most 3 colors. Let $uv$ and $uv'$ be two adjacent edges of $G$. The corresponding demands are assigned different waypoints in our feasible routing, otherwise some edge would be overloaded, so the coloring is proper.
\end{proof}

For the general case, we can strengthen the hardness to vertex cover number~$\tau=2$.
Note that the case~$\tau = 1$ is trivial as the input is a star and waypoints are useless.

\begin{theorem}
\label{thm:binpacking}
    \sr{} is strongly NP-complete on undirected, bidirected and directed graphs even with vertex cover number $\tau=2$, $k=1$ and unit edge/arc weights.
\end{theorem}

\begin{proof}
    We provide a polynomial-time reduction from the strongly NP-complete problem \binpacking~\cite{garey1979computers}.
    \pbdef{\binpacking}{A set of $\ell$ items of sizes $a_1, \ldots, a_\ell \in \mathbb{N}$, a number of bins $b \in \mathbb{N}$ and a bin capacity $C \in \mathbb{N}$.}{Is there a partition of $\lbrace 1, \ldots, \ell \rbrace$ into $b$ disjoint sets $I_1, \ldots, I_b$ such that the sum of the sizes of the items in each $I_j$ is not greater than $C$?}

    Let $\mathcal{I}=(a_1, \ldots, a_\ell, b, C)$ be an instance of \binpacking. We construct in polynomial time an instance $\mathcal{I'}=(G=(V,E), \omega, c, D, k)$ of \sr{} as follows (see \cref{fig:red_BP}):
	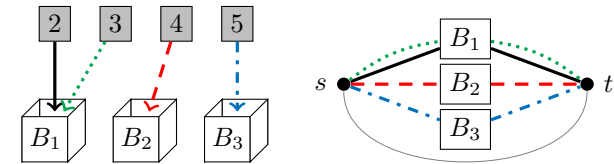
\begin{figure}[t]
		\centering
		\begin{tikzpicture}[scale=0.8]
        \input{figures/red-from-bp}
		\end{tikzpicture}
		\caption{Illustration of the reduction from \binpacking to \sr{} for undirected graph with $k=1$. The number in an item on the left represents its size. The bin capacity is $C=6$. All demands are from $s$ to $t$ in the reduced instance.}
		\label{fig:red_BP}
	\end{figure}

    Set $V = \lbrace B_1, \ldots, B_b \rbrace \cup \lbrace s, t \rbrace$ where vertex $B_j$ correspond to bin $j$, and the two additional vertices $s, t$ will serve as source and target of the demands. For each $j = 1, \ldots, b$, add edges $B_j s$ and $B_j t$ in $E$. Add one last \emph{shortcut} edge $st$ in $E$.
    For all $e \in E$, set $\omega(e) = 1$ and $c(e) = C$.
    For each $i = 1, \ldots, \ell$, create a demand $(s, t, a_i)$ in $D$. Create one last \emph{dummy demand} $(s,t, C)$ in $D$.
    Set $k=1$.
    Clearly, the above transformation is polynomial. 
	We show that $\mathcal{I}$ is a yes-instance if and only if $\mathcal{I'}$ is a yes-instance.
    
    ``$\Rightarrow$'' Let $(I_j)_{j = 1, \ldots, b}$ be a solution of $\mathcal{I}$. For each demand $(s,t,a_i)$, pick as waypoint the $B_j$ such that $i \in I_j$. No waypoint is picked for the dummy demand $(s,t,C)$, so it is routed on $st$. The total flow traversing $s B_j$ and $B_j t$ is equal to the sum of the items in $I_j$ and only the dummy demand is routed on $st$, so we have a feasible routing scheme.
    
    ``$\Leftarrow$'' Conversely, suppose we have a feasible routing scheme for $\mathcal{I'}$. Notice that the dummy demand $(s,t,C)$ either saturates $st$ or it saturates two edges $s B_j, B_jt$ for some $j$. In the latter case, we pick the waypoint $B_j$ for every demand $(s,t,a_i)$ that is routed without waypoint and we route $(s,t,C)$ on $st$ instead. It gives another feasible routing scheme in which every demand $(s,t,a_i)$ has exactly one waypoint among $\lbrace B_1, \ldots, B_b \rbrace$. For each $j=1, \ldots, b$, set $I_j$ as the set of items $i$ such that demand $(s,t,a_i)$ has $B_j$ as waypoint. Then $(I_j)_{j = 1, \ldots, b}$ is a partition of $\lbrace 1, \ldots, \ell \rbrace$ and since no edge capacity is exceeded, the sum of the sizes of the items in $I_j$ is at most $C$.

    Since all flows are from left to right, the reduction is also valid when $G$ is defined as a directed graph with edges directed from $s$ to $\lbrace B_1, \ldots, B_b \rbrace$ to $t$, or as a bidirected graph.
\end{proof}

\citet{JansenKMS13} proved that \unarybinpacking (where the sizes are given in unary encoding) is W[1]-hard parameterized by the number $b$ of bins, that is there is no exact algorithm with running time $f(b) \cdot |\mathcal{I}|_1^{O(1)}$ for any function $f(b)$ (assuming the standard complexity hypothesis FPT $\neq$ W[1]), where $|\mathcal{I}|_1$ is the  size of the unary encoding of the input instance. 
As the reduction provided in the above proof is indeed a parameterized reduction, it follows that there is no exact algorithm for \unarysr with running time $f(n) \cdot |\mathcal{I}'|_1^{O(1)}$ for any instance $\mathcal{I}'$ with $n$ vertices and for any function $f$ (again assuming FPT $\neq$ W[1]), where $|\mathcal{I}'|_1$ is  the size of the  unary encoding of $\mathcal{I}'$.

\begin{corollary}
\label{cor:binpacking}
    \unarysr is  W[1]-hard parameterized by the number of vertices of the graph, even for~$k=1$ and unit weights.
\end{corollary}

\section{Cactus graphs}
\label{sec:cactus_graphs}

\sr{} is trivial on trees: Any pair of vertices is connected by a unique path, so waypoints cannot modify the routes.
In this section, we investigate whether the tractable case can be extended.
While treewidth is well-motivated in practice~\cite{rost2019}, it appears to be unhelpful as the graph constructed in the reduction for \cref{thm:threeedgecol} has treewidth 3.
However, another extension of trees is still manageable for unit capacities and demands: it is the case of cactus graphs. 
We first establish a preliminary result by showing that, on undirected cycles, {\sc Min}~\unitsr, that is, the problem of computing the minimum number of waypoints required to have a feasible solution when there is one, can be solved in polynomial time.

\begin{lemma}
    \label{lem:cycle}
  {\sc Min}~\unitsr is polynomial-time solvable on undirected cycles.
\end{lemma}

\begin{proof}
	Let $C_\ell=(V,E)$ denote a cycle with $\ell \geq 3$ vertices and $D$ a set of demands on $C_\ell$. Notice that in a cycle, the flow function is identical in $FG(s,t)$ and $FG(t,s)$ (only the direction changes) so we consider the forwarding graphs as undirected in this proof.
	Any demand ($s, t) \in D$ such that $s=t$ does not use any edge and can be ignored. 
	We now assume that for every $(s,t) \in D$, $s \neq t$. 
	Two demands \emph{alternate} if their terminals are located in four distinct vertices such that any of the two paths between terminals of one pair contains a terminal of the other pair. Two demands are \emph{duplicates} if $\lbrace s_i, t_i \rbrace = \lbrace s_j, t_j \rbrace$. 
	In the following we differentiate cases according to the number of demands $d=|D|$ (see \cref{fig:cycle} for some illustrations):
    
    If $d=1$ then the demand can be routed without waypoint. 
    
    If $d=2$: If both $FG(s_1, t_1)$ and $FG(s_2, t_2)$ are not ECMP-free, then there is a feasible solution without waypoint. Assume now that at least one of them is ECMP-free.
    \begin{itemize}
        \item If the demands alternate, there is no feasible solution.
        \item If $FG(s_1, t_1)$ and $FG(s_2,t_2)$ are edge-disjoint, there is a feasible solution without waypoint.
        \item If $FG(s_1, t_1) \subsetneq FG(s_2, t_2)$, then demand $(s_2, t_2)$ requires one waypoint to be routed the other way.
        \item If $(s_1, t_1)$ and $(s_2, t_2)$ are duplicates, then one demand must use $FG(s,t)$ without waypoint, and the other must use the rest of the cycle using one waypoint, or two waypoints in the case where $st \in E$ and $\ell$ is even.
    \end{itemize}
    
	If $d \geq 3$: There is a feasible solution if and only if there is no duplicate demand and for each $(s_i, t_i)$, at least one of the two paths from $s_i$ to $t_i$ (clockwise or counterclockwise) does not contain any terminal.
 
    ``$\Rightarrow$'' We prove this direction by contradiction. Assume there are two duplicate demands: according to the case $d=2$, they saturate the cycle, so the third demand cannot be routed. Assume the two paths from $s_1$ to $t_1$ contain at least one terminal each. If we can find suitable $s_2, t_2$, then the demands alternate: either there is no solution to route these two demands alone, or $FG(s_1, t_1)$ and $FG(s_2, t_2)$ are not ECMP-free, saturate the cycle and the third demand cannot be routed. If we can only find two suitable terminals of distinct demands $s_2, s_3$ contained in a path each, it implies that $t_2, t_3$ are in $s_1$ or $t_1$. If $FG(s_2, t_2)$ or $FG(s_3, t_3)$ is not ECMP-free, there is no solution to route these two demands alone. Otherwise, they saturate at least one edge of the path their terminal is in and $(s_1, t_1)$ cannot be routed.

    ``$\Leftarrow$'' If there is no duplicate demand and for each $(s_i, t_i)$, at least one of the two paths from $s_i$ to $t_i$ does not contain any terminal, then either the forwarding graphs are pairwise edge-disjoint and there is a feasible solution without waypoint, or the forwarding graphs are pairwise edge-disjoint except for one $FG(s_i,t_i)$ that contains all the others, and there is a unique feasible solution assigning one waypoint to $(s_i, t_i)$.

    \begin{figure}
        \centering
        \begin{tikzpicture}
        \input{figures/cycles}
        \end{tikzpicture}
        \caption{Illustration of some cycle cases. On the left, a feasible case exploiting ECMP. In the middle, a feasible case requiring two waypoints. On the right, an unfeasible case with three demands.}
        \label{fig:cycle}
    \end{figure}
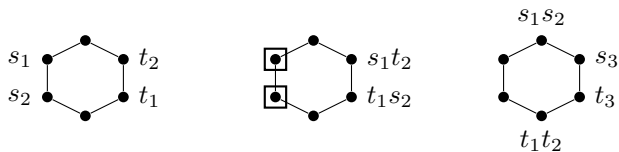

    Only in the last case of $d=2$  there is a waypoint assignment to decide. 
    In all other cases, the minimal solution is \emph{uniquely determined}, either assigning a waypoint to a certain demand or no waypoint at all.
\end{proof}

\begin{theorem}
    \label{thm:cactus}
    \unitsr on undirected cacti can be solved in~$O(kdn)$ time.
\end{theorem}

Let $\mathcal{I}=(G=(V,E), D,k)$ be an instance of \unitsr on  a cactus $G$. 
We use dynamic programming on a tree structure associated with the cactus. 
More precisely, we use the \emph{skeleton} of a cactus, introduced by \citet{burkard1998linear}.

\begin{figure}[t]
    \centering
    \begin{tikzpicture}
    \input{figures/cactus-skeleton}
    \end{tikzpicture}
    \caption{Left: A cactus graph~$G$ where grafts are highlighted by ellipses and hinges are denoted by squares. Right: The skeleton of~$G$ rooted in a block.}
    \label{fig:cactus}
\end{figure}
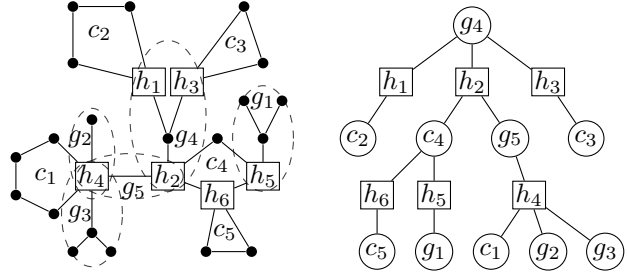

The vertex set $V$ is partitioned into three subsets: a \emph{C-vertex} is a vertex of degree two that is included in exactly one cycle, a \emph{G-vertex} is a vertex not included in any cycle, and the remaining vertices are refereed to as \emph{hinges}. 
A hinge is included in at least one cycle and has degree at least three. 
A \emph{graft} is a maximal subtree of $G$ induced by $G$-vertices and hinges such that a hinge has degree one in a graft. 
A \emph{block} is a cycle or a graft. 
It is easy to see that a cactus consists of blocks attached together by hinges: see \cref{fig:cactus} for an illustration. 

The \emph{skeleton} of $G$ is a tree $S=(V_S, E_S)$ whose $V_S$ represent the blocks and hinges of $G$. 
For distinguishing~$G$ and~$S$, we subsequently refer to~$V$ as \emph{vertex} set and to~$V_S$ as \emph{node} set.
As hinges appear in both~$G$ and~$S$, we refer to them as vertices or nodes, depending on whether we want to emphasize their role in~$G$ or in~$S$.
Let $b \in V_S$ be a block node and let $V(b) \subseteq V$ be the vertices of the corresponding block in $G$. 
There is an edge $bh \in E_S$ between a block node $b$ and a hinge node $h$ if $h \in V(b)$. 
All leaves of $S$ represent blocks and we root $S$ in some block (see \cref{fig:cactus}). 
We root~$S$ in an arbitrary block ($G$ always contains at least one block).

Let $s \in V_S$. 
We denote $S_s$ the subtree of $S$ rooted in $s$.
We say that a demand \emph{visits} a block node if it must use at least one edge of that block in $G$, and that it \emph{visits} a hinge node if it visits two of its adjacent block nodes. 
A demand \emph{leaves} $S_s$ if it visits $s$ and its parent.
Since we consider unit demand weights and unit capacities, we can do the following preprocessing.

\begin{rrule}
    \label{rr:number-demand-leaves} 
    Let~$S_s$ be a subtree of~$S$ rooted in a graft or in a hinge whose parent is a graft.
    If more than one demand leaves~$S_s$, then return that there is no solution.
    
    Let~$S_s$ be a subtree of~$S$ rooted in a cycle or in a hinge whose parent is a cycle.
    If more than two demands leave~$S_s$, then return that there is no solution.
\end{rrule}
We subsequently assume that \cref{rr:number-demand-leaves} is not applicable (otherwise we are done) and perform bottom-up dynamic programming on~$S$.
To this end, we keep in each node the partial solution(s) that minimize the number of waypoints assigned to demands that leave the subtree.
Due to \cref{rr:number-demand-leaves}, there are at most two such demands.

\begin{rrule}
    \label{rr:number-demand-leaves2}
    Let $s \in V_S$. 
    If no demand leaves~$S_s$ and~$s$ is not the root, then we disconnect~$S_s$ from the rest of~$S$ and consider~$S_s$ as its own instance which we can solve separately.
\end{rrule}
We subsequently assume that \cref{rr:number-demand-leaves2} is not applicable, thus the root is the only node without a demand leaving it.
Let $s$ be a node that is not the root. 
If only one demand~$i$ leaves~$S_s$, then the associated partial solution is~$P_s = \{i\}$ and one integer~$T_s^i = x$ indicating the minimum number~$x$ of waypoints assigned to demand~$i$ within~$S_s$ over all possible solutions.
If two demands~$i$ and~$j$ leave~$S_s$, then the associated partial solution is a pair $P_s=\lbrace i,j \rbrace$ and two tables of $k+1$ integers $T_s^{i}$ and $T_s^j$ that correspond to the waypoint assignments within~$S_s$: for each $0 \leq x \leq k$, $T_s^{i}[x]=y$ if $y$ is the minimum number of waypoints assigned to demand~$j$ within $S_s$ when at most~$x$ waypoints are assigned to demand~$i$ within $S_s$.
If there is no valid solution assigning~$x$ waypoints to demand~$i$, then $T_s^{i}[x]= \infty$. 
The table $T_s^j$ is defined analogously.

We subsequently describe how to compute the partial solutions by distinguishing three cases: the root of the current subtree is a graft node, a hinge node, or a cycle node.
For simplicity, we set~$T_s^{i}[x]= \infty$ for~$x < 0$, $x = \infty$, or~$x = -\infty$.
Hence, we omit explicitly handling trivial cases that do not have solutions.

\paragraph{Graft} \label{case:graft}
This is the easiest case: in trees any two vertices are connected via a unique path, thus waypoints are useless.
Let $g$ be a graft node or a graft leaf. 
For each demand visiting~$g$, compute its entry and exit vertices and the unique path connecting them in $G$. 
If any edge has at least two demands flowing on it, then there is no solution.
Let~$(s_i, t_i)$ be a demand that visits~$g$ but does not leave~$S_g$. Depending on whether $s_i, t_i \in V(g)$, there are at most two child hinge nodes~$h_1$ and~$h_2$ of~$g$ such that~$P_{h_1}=P_{h_2}= \lbrace i \rbrace$.
Compute~$T_{h_1}^i + T_{h_2}^i$, taking the corresponding term to zero if a child does not exist. If~$T_{h_1}^i + T_{h_2}^i > k$, including when some term is $\infty$, then there is no solution.
In a case without solution, if~$g$ is the root of~$S$, return false, or if $g$ is not the root, set the partial solution: $P_g = \lbrace j \rbrace$ and~$T_g^j = \infty$ (where demand~$j$ leaves~$g$). 
Otherwise, return true if~$g$ is the root of~$S$, or set the partial solution: $P_g = \lbrace j \rbrace$ and if there is a child hinge~$h$ with~$P_h=\{j\}$, then set~$T_g^j = T_h^j$, otherwise set $T_g^j = 0$.

\paragraph{Hinge}
Hinges are more complicated than grafts, as up to two demands can leave it and its child nodes can also have two demands leaving into the hinge. Note that a hinge is never a root or leaf of $S$.
This introduces dependencies between the demands: As seen in the proof of \cref{lem:cycle}, there can be a choice of assigning the waypoint(s) to either demand visiting a cycle.

To capture these dependencies, we define the \emph{dependency multigraph} $H(s)$ of a node~$s \in V_S$, which we later also use for cycle nodes: 

The vertices are the parent~$p$ of $s$ (except when $s$ is the root) and the children~$c_1, \ldots, c_\ell$ of $s$;
    there is an edge $c_i c_j$ between two children for each element of $P_{c_i} \cap P_{c_j}$;
    there is an edge $c_i p$ between a child and the parent for each demand of $P_{c_i}$ that visits $p$.
\begin{observation}
	\label{obs:H(s)-max-deg-2}
    If \cref{rr:number-demand-leaves,rr:number-demand-leaves2} are not applicable, then $H(s)$ has maximum degree 2 and its connected components are paths or cycles (no isolated vertex).
\end{observation}

Let $h$ be a hinge. To compute the partial solutions for~$h$, we first compute $H(h)$. 
Observe that each connected component of $H(h)$ can be processed independently.
Thus, for each connected component~$C$ of $H(h)$ that does not contain~$p$, we create a new cactus graph (also rooted in a block) that we solve independently.
If any of these graphs does not have a solution, we set all table entries for~$h$ to $\infty$.
Thus, $H(h)$ is a path or a cycle containing $p$.

Assume that $H(h)$ is a path. 
Our plan is to repeatedly process the endpoints of the path until only~$p$ remains:
At least one of the endpoints is a child $c$. 
Since $c$ has degree one in $H(h)$, we have $P_c=\lbrace i \rbrace$ for some~$i$ and~$T_c^i$ waypoints are required for demand~$i$ within~$S_c$.
Denote~$c'$ the neighbor of~$c$ in~$H(h)$ and~$P_{c'} = \{i,j\}$ (assume~$c' \ne p$).
Then at most $k - T_c^i$ waypoints can be used for demand~$i$ in~$S_{c'}$ in any feasible solution.
Hence~$T_{c'}^i[k - T_c^i]$ denotes the number of waypoints demand~$j$ requires within~$S_{c'}$; allowing us to continue with the neighbor of~$c'$ in the path.
Following this procedure, we can process the path node by node.
If at any point the number of required waypoints is $\infty$, then there is no solution; recall we set~$T_s^{i}[-\infty]=T_s^{i}[\infty]=\infty$. 
When only~$p$ remains, then the above procedure computes the minimum number of waypoints that the leaving demand(s) require within~$H_h$; denote these number(s) by~$k_i$ (one demand~$i$ leaving~$h$) or by~$k_i, k_j$ (two demands $i$ and~$j$ leaving~$h$).
If one demand~$i$ leaves~$h$, then set~$P_h= \lbrace i \rbrace, T_h^i = k_i$.
If two demands~$i$ and~$j$ leave~$h$, then set~$P_h=\lbrace i,j \rbrace$, $T_h^i[x] = k_j$ for all~$x \ge k_i$ and~$T_h^j[x] = k_i$ for all~$x \ge k_j$.

It remains to consider the case~$H(h)$ is a cycle. Assume it is a cycle of length 2 containing $p$ and a child $c$.
This implies that the two demands of $P_c$ leave $S_h$. 
Then copy the partial solution of $c$ to $h$.

Assume it is a longer cycle containing $p$. It implies that two demands leave $S_h$.
Denote $P_h=\lbrace i,j \rbrace$. 
We describe how to fill $T_h^{i}$ and the same applies to $T_h^{j}$. 
For each value $0 \leq x \leq k$, we compute $y=T_h^{i}[x]$ by reducing it to the path-case above:
Since~$h$ is a hinge, no waypoint needs to be set on~$h$.
Let~$c$ be the child of~$h$ with $i \in P_{c}$. 
By definition of the table~$T_h^{i}[x]$, at most~$x$ waypoints should be used for demand~$i$ in~$S_h$.
Hence, $T_c^{i}[x]$ waypoints are required within~$S_c$ for the second demand leaving~$S_c$.
We can thus pretend that the edge~$cp$ in~$H(h)$ is deleted and repeat the procedure used in paths above to compute the number~$y$ of waypoints needed on demand~$j$ within $S_h$.

\paragraph{Cycle}
Let $c$ be a cycle node. 
Similarly to the graft case, for each demand visiting $c$, compute its entry and exit vertices; this leaves us with a cycle and some demands on it that can be solved using \cref{lem:cycle}. If $c$ is the root and a leaf, then the problem reduces to the cycle-instance.
If this cycle-instance has no solution, return false if $c$ is the root, otherwise initialize the partial solution for~$c$ and fill the table entrie(s) with $\infty$.

Assume that the cycle-instance has a unique minimal solution without waypoints.
If $c$ is a leaf, initialize the partial solution accordingly, filling the table entrie(s) with zeros. 
If $c$ is an inner node, compute $H(c)$: its connected components can be processed independently. If any of them returns false, fill the partial solution of $c$ with $\infty$.
To adapt the previous procedure to a component without the parent $p$, pick a vertex to act as $p$. 
For a path of $H(c)$, pick an arbitrary endpoint $h$ with partial solution $P_h = \lbrace i \rbrace, T_h^{i}=x$. 
Compute the minimum number~$x'$ of waypoints to assign to demand $i$ outside of $S_h$ according to the previous procedure, and return true if $x+x'\leq k$ and false otherwise.
Similarly, for a cycle of length 2 of $H(c)$, pass the partial solution. For a longer cycle, pick an arbitrary child $h$ with partial solution $P_h= \lbrace i,j \rbrace$, $T_h^{i}, T_h^j$. Compute ${T_h^{i}}'$ and ${T_h^j}'$ the waypoint assignments elsewhere than $S_h$ according to the previous procedure, and then return true if and only if there exists $x$ such that $T_h^{i}[x]=y$ and ${T_h^{j}}'[k-y]=x' \leq k-x$. 
The component with $p$ can be processed as for a hinge, but taking into account the demand(s) that may leave $S_c$ and have a terminal in $V(c)$. If $c$ is the root, there are only components without $p$.

Assume that the cycle-instance has a unique minimal solution assigning one waypoint to demand $i$. Proceed identically as in the previous case, but taking into account that additional waypoint. If $c$ is a leaf and $i \in P_c$, its waypoint assignments are 1. If $c$ is an inner node or the root, and demand $i$ corresponds to an edge in $H(c)$ that is not incident to $p$, take into account that the limit is $k-1$ for that demand in the computation. If $i \in P_c$, incorporate that waypoint in the partial solution of $c$ after computing it similarly as before.

Assume that the cycle-instance has two minimal solutions.
Note that the proof of \cref{lem:cycle} shows that it can only happen when two demands $i$ and $j$ visit the cycle with the same entry and exit, and they saturate the cycle. Thus $H(c)$ has two edges and $P_c \subseteq \lbrace i, j \rbrace$. 
Simply try the two solutions with the same method as for a unique minimal solution, and take the better result. If only one demand leaves, then the better solution is the one minimizing the number of waypoints assigned to the demand in $P_c$. If both leaves, the partial solution contains the two solutions. If none leaves, then $c$ is the root and any of the two works.

This finishes the description of the dynamic program.
\begin{proof}[Proof of \cref{thm:cactus}]
	Compute the skeleton $S$ of $G$. 
	The dynamic programming algorithm solves the problem from the leaves of $S$ to its root. 
	For each node of $S$, we apply the respective case \emph{a)} to \emph{c)} to compute the table entries.
	A solution can be reconstructed by backtracking on the filled tables.
	
	For each node~$s$ we compute at most~$2(k+1)$ entries to store in the table.
	Let~$n_s$ denote the number of vertices in~$V(s)$ and~$d_s$ the number of demands with one endpoint in~$S_s$.
	The size of the dependency multigraph~$H(s)$ is at most $d_s$ because every edge corresponds to a distinct demand and the computations along the paths and cycles in~$H(s)$ are simple arithmetic operations.
	Thus, each table entry can be computed in~$O(d_s n_s)$ time, which leads to~$O(k d_s n_s)$ time per node.
	Summing up over all nodes yields~$O(kdn)$ time.
\end{proof}

We conclude by establishing the NP-hardness for the general case even when~${k=1}$, that is, unit capacities and demands are essential in the above algorithm.
Thus, generalizing its ideas, one should aim at approximation or parameterized algorithms.

\begin{theorem}
    \sr{} is strongly NP-complete on undirected and bidirected cacti even for $k=1$ and unit edge/arc weights. 
\end{theorem}

\begin{proof}
    We provide a reduction from the strongly NP-complete problem \threepart \cite{garey1979computers} to \sr{} on an undirected cactus.

    \pbdef{\threepart}{A multiset $A$ of $3\ell$ positive integers and a bound $B \in \mathbb{N}$ such that $\sum_{a \in A} a = \ell B$ and $\forall a \in A$, $\frac{B}{4} < a < \frac{B}{2}$.}{Is there a partition of $A$ into $\ell$ sets $A_1,\ldots, A_\ell$ containing 3 integers each and such that for each $j =1, \ldots, \ell$, $\sum_{a \in A_j} a = B$?}
    
    Let $\mathcal{I}=(A, B)$ be an instance of \threepart. We construct in polynomial time an instance $\mathcal{I}'=(G=(V,E), \omega, c, D, k)$ of \sr{} where $G$ is a cactus as follows (see \cref{fig:red_BP2}):
    Start with an empty graph. For each $j=1, \ldots, \ell$, create a triangle graph $T_j$. 
	Connect the triangles in a chain with one edge between them (see \cref{fig:red_BP2}).
	Denote the left-most vertex~$s$ and the right-most vertex~$t$.
    For each edge $e \in E$, set $\omega(e)=1$. For each triangle~$T_j$, set the capacity of the top edge to $(\ell-1)B$ and the capacity of the two bottom edges to $B$. 
    Set the capacity of all other edges to $\ell B$.
    For each element $a \in A$, create a demand $(s,t,a)$ in $D$.
    Set $k=1$.
    Clearly the above transformation is polynomial. 
    We show that $\mathcal{I}$ is a yes-instance if and only if $\mathcal{I}'$ is a yes-instance.

    ``$\Rightarrow$'' Suppose that there is a solution $(A_j)_{j = 1, \ldots, \ell}$ to $\mathcal{I}$. For each demand $(s,t,a)$, pick as waypoint the bottom vertex of the triangle $T_j$ such that $a \in A_j$. 
    The segment path of $(s, t, a)$ is the straight line through all triangles except for $T_j$ where it traverses the bottom vertex. The load on the two bottom edges of each $T_j$ is the sum of elements in $A_j$ which is $B$. The load of the top edge of $T_j$ is the sum of elements not in $A_j$ which is exactly $(\ell-1) B$, so no edge is overloaded.

    ``$\Leftarrow$'' Suppose there is a feasible routing scheme for $\mathcal{I}'$. It is easy to see that in any feasible routing scheme, all edges are saturated and each demand has exactly one waypoint among the bottom vertices of the triangles. Let $w_j$ be the bottom vertex of $T_j$. Since $\frac{B}{4} < a < \frac{B}{2} \quad \forall a \in A$ and the edges incident to $w_j$ have capacity $B$, exactly 3 demands have $w_j$ as waypoint. Let $A_j = \lbrace a \in A : (s,t,a)$ has $w_j$ as waypoint$\rbrace$. Then $(A_1, \ldots, A_\ell)$ is a solution of $\mathcal{I}$.
    
    The same arguments apply if the cactus is bidirected.
    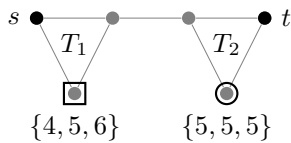
\begin{figure}[t]
    \centering
    \begin{tikzpicture}
    \input{figures/red-from-3partition}
    \end{tikzpicture}
    \caption{Illustration of the reduction from \threepart to \sr{} on a cactus. The starting instance is $\lbrace 4,5,5,5,5,6 \rbrace$ with $\ell=2$ and $B=15$. All demands are from $s$ to $t$ and below the waypoints are indicated the bandwidth requirement of the demands using it.}
    \label{fig:red_BP2}
    \end{figure}
\end{proof}

\section{Conclusion And Future Works}

We provide strong intractability results for \sr{}. 
While these show the limitations on what is (probably) algorithmically feasible, our polynomial-time algorithm for \unitsr{} gives rise for future work.
The reduction from the number problem \textsc{3-Partition} that shows strongly NP-hardness of \sr{} does not yield approximation lower bounds.
Indeed, an interesting question is whether known approximation algorithms for \textsc{Bin Packing}~\cite{CCGMV13} (to deal with the demands and capacities) can be combined with the insights of our polynomial-time algorithm to get an approximation algorithm for the variant with non-unit demands and capacities.

Another research direction is to incorporate the structure of the demand graph (which we ignored) in the analysis:
Typical Internet Service Provider (ISP) networks have a hierarchical structure, with nodes divided into access nodes and backbone nodes. 
Access nodes handle the incoming and outgoing traffic and are grouped into so-called Point of Presence (PoP). 
Each PoP is connected to the rest of the network through a few backbone nodes (usually at least two to ensure robustness against link/node failures), forming a star-like structure \cite{quoitin2009igen} (see, for instance, ``brain'' network from SNDLib~\cite{sndlib}). 
In such a topology, the demand graph induced only by the backbone nodes is almost complete.
Consequently, it is natural to ask for a FPT algorithm with respect, for instance, to the combined parameter \emph{treewidth} of the network and \emph{distance to cluster} of the demand graph (\emph{i.\,e.} the size of a smallest vertex subset whose deletion makes G a collection of disjoint cliques). 

\printbibliography

\end{document}

%% file: figures/intro-example.tex
        \node[black vertex] (1) at (1,-1) [label=above:$a$] {};
        \node[gray vertex] (2) at (2,0) [label=above:$b$] {};
        \node[gray vertex] (3) at (2,-1) [label=above:$c$] {};
        \node[gray vertex] (4) at (3,0) [label=above:$d$] {};
        \node[gray vertex] (5) at (3,-1) [label=above:$e$] {};
        \node[gray vertex] (6) at (3,-2) [label=below:$f$] {};
        \node[black vertex] (7) at (4,-1) [label=above:$g$] {};
        
        \draw[black arc] (1) -- (2) node [midway, fill=white,circle,inner sep=0pt] {$\nicefrac{1}{2}$};
        \draw[black arc] (1) -- (3) node [midway, fill=white,circle,inner sep=0pt] {$\nicefrac{1}{2}$};
        \draw[black arc] (2) -- (4) node [midway, fill=white,circle,inner sep=0pt] {$\nicefrac{1}{2}$};
        \draw[black arc] (3) -- (5) node [midway, fill=white,circle,inner sep=0pt] {$\nicefrac{1}{4}$};
        \draw[black arc] (3) -- (6) node [midway, fill=white,circle,inner sep=0pt] {$\nicefrac{1}{4}$};
        \draw[black arc] (4) -- (7) node [midway, fill=white,circle,inner sep=0pt] {$\nicefrac{1}{2}$};
        \draw[black arc] (5) -- (7) node [midway, fill=white,circle,inner sep=0pt] {$\nicefrac{1}{4}$};
        \draw[black arc] (6) -- (7) node [midway, fill=white,circle,inner sep=0pt] {$\nicefrac{1}{4}$};
    
    \begin{scope}[xshift=3.75cm]
        \node[black vertex] (1) at (1,-1) [label=above:$a$] {};
        \node[gray vertex] (2) at (2,0) [label=above:$b$] {};
        \node[gray vertex] (3) at (2,-1) [label=above:$c$] {};
        \node[gray vertex] (4) at (3,0) [label=above:$d$] {};
        \node[gray vertex] (5) at (3,-1) [label=above:$e$] {};
        \node[gray vertex] (6) at (3,-2) [label=below:$f$] {};
        \node[black vertex] (7) at (4,-1) [label=above:$g$] {};
        \node[square waypoint] at (2,-1) {};
        
        \draw[gray edge] (1) -- (2);
        \draw[black arc] (1) -- (3) node [midway, fill=white,circle,inner sep=0pt] {$1$};
        \draw[gray edge] (2) -- (4);
        \draw[black arc] (3) -- (5) node [midway, fill=white,circle,inner sep=0pt] {$\nicefrac{1}{2}$};
        \draw[black arc] (3) -- (6) node [midway, fill=white,circle,inner sep=0pt] {$\nicefrac{1}{2}$};
        \draw[gray edge] (4) -- (7);
        \draw[black arc] (5) -- (7) node [midway, fill=white,circle,inner sep=0pt] {$\nicefrac{1}{2}$};
        \draw[black arc] (6) -- (7) node [midway, fill=white,circle,inner sep=0pt] {$\nicefrac{1}{2}$};
    \end{scope}

%% file: figures/tricky-example.tex
        \node[small black vertex] (t1) at (0,0) [label=left:$s_2 s_3$] {};
        \node[small black vertex] (t2) at (0.5,1) {};
        \node[small black vertex] (t3) at (1,0) [label=below:$t_3$] {};
        \node[small black vertex] (t4) at (1.5,1) {};
        \node[small black vertex] (t5) at (2,0) [label=below left:$v$] {};
        \node[small black vertex] (b1) at (3,0) {};
        \node[small black vertex] (b2) at (4,0) {};
        \node[small black vertex] (b3) at (5,0) {};
        \node[small black vertex] (b4) at (6,0) {};
        \node[small black vertex] (b5) at (7,0) [label=right:$t_2$] {};
        \node[small black vertex] (a1) at (3.5,1) [label=above:$t_1$] {};
        \node[small black vertex] (a2) at (5.5,1) {};
        \node[small black vertex] (c1) at (3.5,-1) [label=below:$s_1$] {};
        \node[small black vertex] (c2) at (5.5,-1) {};
        \draw[black edge] (t1)--(t2) (t3)--(t1);
        \draw[black edge] (t2)--(t4)--(t5)--(t3);
        \draw[black edge] (t5)--(a1)--(a2)--(b5);
        \draw[black edge] (t5)--(b1)--(b2)--(b3)--(b4)--(b5);
        \draw[black edge] (t5)--(c1)--(c2)--(b5);
        \node[square waypoint, red] at (b3) {};
        \draw[dashed arc, red] (c1) to[bend right=15] (t5);
        \draw[dashed arc, red] (t5) to[bend right=15] (b1);
        \draw[dashed arc, red] (b1) to[bend right=15] (b2);
        \draw[dashed arc, red] (b2) to[bend right=15] (b3);
        \draw[dashed arc, red] (c1) to[bend left=15] (c2);
        \draw[dashed arc, red] (c2) to[bend left=15] (b5);
        \draw[dashed arc, red] (b5) to[bend left=15] (b4);
        \draw[dashed arc, red] (b4) to[bend left=15] (b3);
        \draw[dashed arc, red] (b3) to[bend left=15] (b4);
        \draw[dashed arc, red] (b4) to[bend left=15] (b5);
        \draw[dashed arc, red] (b5) to[bend left=15] (a2);
        \draw[dashed arc, red] (a2) to[bend left=15] (a1);
        \draw[dashed arc, red] (b3) to[bend right=15] (b2);
        \draw[dashed arc, red] (b2) to[bend right=15] (b1);
        \draw[dashed arc, red] (b1) to[bend right=15] (t5);
        \draw[dashed arc, red] (t5) to[bend right=15] (a1);
        \node[round waypoint, RoyalBlue] at (t4) {};
        \draw[dashdotted arc, RoyalBlue] (t1) to[bend left=15] (t2);
        \draw[dashdotted arc, RoyalBlue] (t2) to[bend left=15] (t4);
        \draw[dashdotted arc, RoyalBlue] (t4) to[bend left=15] (t5);
        \draw[dashdotted arc, RoyalBlue] (t5) to[bend left=15] (a1);
        \draw[dashdotted arc, RoyalBlue] (a1) to[bend left=15] (a2);
        \draw[dashdotted arc, RoyalBlue] (a2) to[bend left=15] (b5);
        \draw[dashdotted arc, RoyalBlue] (t5) to[bend right=15] (c1);
        \draw[dashdotted arc, RoyalBlue] (c1) to[bend right=15] (c2);
        \draw[dashdotted arc, RoyalBlue] (c2) to[bend right=15] (b5);
        \draw[dotted arc, Green] (t1) to[bend right=15] (t3);

%% file: figures/triangle-chain.tex
        \foreach \i in {1,2,3,4}{
            \begin{scope}[xshift= \i cm]
                \node[small gray vertex] (a\i) at (0,1) {};
                \node[small gray vertex] (c\i) at (0.5,0) {};
                \node[small gray vertex] (b\i) at (1,1) {};
                \draw[gray edge] (a\i) -- (b\i) -- (c\i) --(a\i);
            \end{scope}
        }
        \node[black vertex] (s1) at (-1, 1) [label=left:$s_1$] {};
        \node[black vertex] (s2) at (-1, 0) [label=left:$s_2$] {};
        \draw[dashed edge, red] (s1) -- (a1) node [midway, fill=white] {$\cdots$};
        \draw[dotted edge, Green] (s2) -- (a1) node [midway, sloped, fill=white] {$\cdots$};
        \node[square waypoint, red] (w1) at (c1) {};
        \node[square waypoint, red] (w2) at (c2) {};
        \node[round waypoint, Green] (w3) at (c3) {};
        \node[round waypoint, Green] (w4) at (c4) {};
        \draw[dashed edge, red] (a1) to[bend right=15] (c1);
        \draw[dashed edge, red] (c1) to[bend right=15] (b1);
        \draw[dashed edge, red] (a2) to[bend right=15] (c2);
        \draw[dashed edge, red] (c2) to[bend right=15] (b2);
        \draw[dashed edge, red] (a3) to[bend left=15] (b3);
        \draw[dashed edge, Green] (a4) to[bend left=15] (b4);
        \draw[dotted edge, Green] (a1) to[bend left=15] (b1);
        \draw[dotted edge, Green] (a2) to[bend left=15] (b2);
        \draw[dotted edge, Green] (a3) to[bend right=15] (c3);
        \draw[dotted edge, Green] (c3) to[bend right=15] (b3);
        \draw[dotted edge, Green] (a4) to[bend right=15] (c4);
        \draw[dotted edge, Green] (c4) to[bend right=15] (b4);
        \node[black vertex] (t1) at (7,1) [label=right:$t_1$] {};
        \node[black vertex] (t2) at (7,0) [label=right:$t_2$] {};
        \draw[dashed edge, red] (b4) -- (t1) node [midway, fill=white] {$\cdots$};
        \draw[dotted edge, Green] (b4) -- (t2) node [midway, sloped, fill=white] {$\cdots$};

%% file: figures/red-from-2edp.tex
		\node[black vertex] (s3) at (0,2) [label=left:$s_3$] {};
		\node[black vertex] (t3) at (0,0) [label=left:$t_3$] {};
		\draw[black arc] (s3) -- (t3);
		\node[gray vertex] (s1) at (4,2) [label=right:$s_1$] {};
		\node[gray vertex] (s2) at (5.5,2) [label=right:$s_2$] {};
		\node[gray vertex] (t1) at (3,0) [label=right:$t_1$] {};
		\node[gray vertex] (t2) at (5,0) [label=right:$t_2$] {};
		\node[gray vertex] (u1) at (3.5,1.5) [label=right:$u$] {};
		\node[gray vertex] (u2) at (3.5,0.5) [label=right:$v$] {};
		\node[small gray vertex] (u12) at (3.5,1) {};
		\draw[rounded corners, color=gray] (2.5,-0.5) rectangle (6.5, 2.5) {};
		\node at (4.5, -1) {\textcolor{gray}{$S_2(G)$}};

		\draw[black arc] (s1) to[bend right=55] (s3);
		\draw[black arc] (s2) to[bend right=75] (s3);
		\draw[black arc] (u1) to[bend right=40] (s3);
		\draw[black arc] (u2) to[bend right=15] (s3);
		\draw[black arc] (t1) to[bend left=25] (s3);
		\draw[black arc] (t2) to[bend left=15] (s3);
		\draw[black arc] (t3) to[bend right=15] (t1);
		\draw[black arc] (t3) to[bend right = 30] (t2);
		\draw[gray arc] (u1)--(u12);
		\draw[gray arc] (u12)--(u2);
		\begin{scope}[xshift=7cm]
			\node[gray vertex] (s1') at (3,2) [label=right:$s_1$] {};
			\node[gray vertex] (s2') at (4.5,2) [label=right:$s_2$] {};
			\node[gray vertex] (t1') at (2,0) [label=right:$t_1$] {};
			\node[gray vertex] (t2') at (4,0) [label=right:$t_2$] {};
			\node[black vertex] (s3') at (1,2) [label=left:$s_3$] {};
			\node[black vertex] (t3') at (1,0) [label=left:$t_3$] {};
			\draw[rounded corners, color=gray] (1.5,-0.5) rectangle (5.5, 2.5) {};
			\node at (3.5, -1) {\textcolor{gray}{$S_2(G)$}};
			\draw[black arc] (s1') -- (t1');
			\draw[black arc] (s2') -- (t2');
			\draw[black arc] (s3') -- (t3');
		\end{scope}

%% file: figures/red-from-2d1sp1.tex
\node[black vertex] (s3) at (0,6) [label=left:$s_3$] {};
		\node[black vertex] (s4) at (0,5) [label=left:$s_4$] {};
		\node[small gray vertex] (1) at (1,6) {};
		\node[small gray vertex] (2) at (1.25,5.5) {};
		\node[small gray vertex] (3) at (1.5,6) {};
		\draw[gray edge] (1) -- (2) -- (3) -- (1);
		\node at (2.5, 6) {\textcolor{gray}{$\cdots$}};
		\node[small gray vertex] (4) at (3.5,6) {};
		\node[small gray vertex] (5) at (3.75,5.5) {};
		\node[small gray vertex] (6) at (4,6) {};
		\draw[gray edge] (4) -- (5) -- (6) -- (4);
		\draw[rounded corners, color=gray] (0.5,5) rectangle (4.5, 6.5) {};
		\node at (2.5,7) {\textcolor{gray}{$\nabla_{2k}$}};
		\draw[black edge] (s3)--(1);
		\draw[black edge] (s4)--(1);
		\node[black vertex] (s3') at (5,6) [label=above:$s_3'$] {};
		\node[black vertex] (s4') at (5,5) [label=right:$s_4'$] {};
		\draw[black edge] (6)--(s3');
		\draw[black edge] (6)--(s4');
		\node[small black vertex] (7) at (6,6) {};
		\node[small black vertex] (8) at (7,6) {};
		\node[small black vertex] (9) at (8,6) {};
		\node[small black vertex] (10) at (9,6) {};
		\node[small black vertex] (11) at (10.5,6) {};
		\node[small black vertex] (12) at (11.5,6) {};
		\draw[black edge] (s3')--(7)--(8)--(9)--(10);
		\draw[black edge] (10)--(11) node [midway, fill=white] {$\cdots$};
		\node[black vertex] (t3) at (12.5,6) [label=right:$t_3$] {};
		\draw[black edge] (11)--(12)--(t3);
		\draw[snake=brace] (5.75,6.25) -- (11.75,6.25);
		\node at (8.5,6.8) {$n$ vertices};
		\node[gray vertex] (t2) at (7,5) [label=right:$t_2$] {};
		\node[gray vertex] (t1) at (7,3.5) [label=right:$t_1$] {};
		\node[gray vertex] (s2) at (11,5) [label=left:$s_2$] {};
		\node[gray vertex] (s1) at (11,3.5) [label=left:$s_1$] {};
		\node[small gray vertex] (u) at (8,4.5) [label=below:$u$] {};
		\node[small gray vertex] (v) at (10,4.5) [label=below:$v$] {};
		\node[small gray vertex] (uv1) at (8.5,4.5) {};
		\node[small gray vertex] (uv2) at (9.5,4.5) {};
		\draw[gray edge] (u)--(uv1);
		\draw[gray edge] (uv2)--(v);
		\draw[gray edge] (uv1)--(uv2) node [midway, fill=white] {$\cdots$};
		\draw[rounded corners, color=gray] (6.5,3) rectangle (11.5, 5.5) {};
		\node at (9,2.5) {\textcolor{gray}{$S_{n+2}(G)$}};
		\draw[black edge] (t1) to[bend left=15] (7);
		\draw[black edge] (t2) -- (8);
		\draw[black edge] (u) -- (9);
		\draw[black edge] (v) -- (10);
		\draw[black edge] (s2) -- (11);
		\draw[black edge] (s1) to[bend right=20] (12);
		\node[black vertex] (t4) at (0,4) [label=left:$t_4$] {};
		\node[black vertex] (t1') at (0,3) [label=left:$t_1'$] {};
		\node[small gray vertex] (13) at (1,4) {};
		\node[small gray vertex] (14) at (1.25,3.5) {};
		\node[small gray vertex] (15) at (1.5,4) {};
		\draw[gray edge] (13) -- (14) -- (15) -- (13);
		\node at (2.5,4) {\textcolor{gray}{$\cdots$}};
		\node[small gray vertex] (16) at (3.5,4) {};
		\node[small gray vertex] (17) at (3.75,3.5) {};
		\node[small gray vertex] (18) at (4,4) {};
		\draw[gray edge] (16) -- (17) -- (18) -- (16);
		\draw[rounded corners, color=gray] (0.5,3) rectangle (4.5,4.5) {};
		\node at (2.5,2.5) {\textcolor{gray}{$\nabla_{k-d_G(s_1,t_1)}$}};
		\draw[black edge] (s4') -- (18);
		\draw[black edge] (t1) -- (18);
		\draw[black edge] (t4) -- (13);
		\draw[black edge] (t1') -- (13);

%% file: figures/red-from-2d1sp2.tex
\node[black vertex] (s3) at (0,6) [label=left:$s_3$] {};
		\node[black vertex] (s4) at (0,5) [label=left:$s_4$] {};
		\draw[rounded corners, color=gray] (0.5,5) rectangle (4.5, 6.5) {};
		\node at (2.5,7) {\textcolor{gray}{$\nabla_{2k}$}};
		\node[black vertex] (s3') at (5,6) [label=above:$s_3'$] {};
		\node[black vertex] (s4') at (5,5) [label=right:$s_4'$] {};
		\node[small black vertex] (7) at (6,6) {};
		\node[small black vertex] (12) at (11.5,6) {};
		\draw[draw=none] (7) -- (12) node [midway] {$\cdots$};
		\node[black vertex] (t3) at (12.5,6) [label=right:$t_3$] {};
		\draw[snake=brace] (5.75,6.25) -- (11.75,6.25);
		\node at (8.5,6.8) {$n$ vertices};
		\node[gray vertex] (t2) at (7,5) [label=right:$t_2$] {};
		\node[gray vertex] (t1) at (7,3.5) [label=right:$t_1$] {};
		\node[gray vertex] (s2) at (11,5) [label=left:$s_2$] {};
		\node[gray vertex] (s1) at (11,3.5) [label=above:$s_1$] {};
		\draw[rounded corners, color=gray] (6.5,3) rectangle (11.5, 5.5) {};
		\node at (9,2.5) {\textcolor{gray}{$S_{n+2}(G)$}};
		\node[black vertex] (t4) at (0,4) [label=left:$t_4$] {};
		\node[black vertex] (t1') at (0,3) [label=left:$t_1'$] {};
		\draw[rounded corners, color=gray] (0.5,3) rectangle (4.5,4.5) {};
		\node at (2.5,2.5) {\textcolor{gray}{$\nabla_{k-d_G(s_1,t_1)}$}};
		\draw[black arc] (s1) to[bend right=10]  (t1');
		\draw[black arc] (s2)to[bend left=40] (t2);
		\draw[black arc] (s3) to[bend right=10]  (t3);
		\draw[black arc] (s4)-- (t4);

%% file: figures/red-from-mcc-layout.tex
			\node[black vertex] (s1) at (0,6) [label=left:$s_1$] {};
			\node[black vertex] (s2) at (0,4) [label=left:$s_2$] {};
			\node[black vertex] (sd) at (0,1.5) [label=left:$s_d$] {};
			\node[black vertex] (sd1) at (1.5,7.5) [label=above:$s_{d+1}$] {};
			\node[black vertex] (sd2) at (3.5,7.5) [label=above:$s_{d+2}$] {};
			\node[black vertex] (sdd) at (6,7.5) [label=above:$s_{2d}$] {};
			\node[black vertex] (t1) at (7.5,6) [label=right:$t_1$] {};
			\node[black vertex] (t2) at (7.5,4) [label=right:$t_2$] {};
			\node[black vertex] (td) at (7.5,1.5) [label=right:$t_d$] {};
			\node[black vertex] (td1) at (1.5,0) [label=below:$t_{d+1}$] {};
			\node[black vertex] (td2) at (3.5,0) [label=below:$t_{d+2}$] {};
			\node[black vertex] (tdd) at (6,0) [label=below:$t_{2d}$] {};

			\foreach \i / \k in {1/1, 3/2, 5.5/d}{
				\foreach \j / \l in {1/d, 3.5/2, 5.5/1}{
					\draw[color=gray] (\i,\j) rectangle (\i+1,\j+1) node[pos=.5] {$(\k,\l)$};
				}
			}
			\foreach \j in {6, 4, 1.5}{
				\draw[dashed edge] (0,\j) to (1,\j+0.4);
				\draw[dashed edge] (0,\j) to (1,\j+0.2);
				\draw[dashed edge] (0,\j) to (1,\j-0.4);
				\draw[dashed edge] (7.5,\j) to (6.5,\j+0.4);
				\draw[dashed edge] (7.5,\j) to (6.5,\j+0.2);
				\draw[dashed edge] (7.5,\j) to (6.5,\j-0.4);
			}
			\foreach \i in {1.5, 3.5, 6}{
				\draw[dashed edge] (\i,7.5) to (\i-0.4,6.5);
				\draw[dashed edge] (\i,7.5) to (\i-0.2,6.5);
				\draw[dashed edge] (\i,7.5) to (\i+0.4,6.5);
				\draw[dashed edge] (\i,0) to (\i-0.4,1);
				\draw[dashed edge] (\i,0) to (\i-0.2,1);
				\draw[dashed edge] (\i,0) to (\i+0.4,1);
			}

			\foreach \x in {1,3,5.5}{
				\begin{scope}[xshift=\x cm]
					\draw[dotted edge] (0.1,5.5) to (0.1,4.5);
					\draw[dotted edge] (0.3,5.5) to (0.3,4.5);
					\draw[dotted edge] (0.9,5.5) to (0.9,4.5);
				\end{scope}
			}
			\foreach \y in {1,3.5,5.5}{
				\begin{scope}[yshift=\y cm]
					\draw[dotted edge] (2,0.9) to (3,0.9);
					\draw[dotted edge] (2,0.7) to (3,0.7);
					\draw[dotted edge] (2,0.1) to (3,0.1);
				\end{scope}
			}

			\node at (1.5,2.75) {$\vdots$};
			\node at (3.5,2.75) {$\vdots$};
			\node at (6,2.75) {$\vdots$};
			\node at (4.75,1.5) {$\cdots$};
			\node at (4.75,4) {$\cdots$};
			\node at (4.75,6) {$\cdots$};

%% file: figures/red-from-mcc-intersections.tex
        \begin{scope}[shift={(-5 cm, -0 cm)}]
            \draw[very thick, color=Red] (1,6) -- (1,4);
            \draw[very thick, color=RoyalBlue] (0,5) -- (2,5);
			\node[small black vertex] at (0,5) [label=left:$P_v$] {};
			\node[small black vertex] at (2,5) {};
			\node[small black vertex] at (1,6) [label=above:$Q_u$] {};
			\node[small black vertex] at (1,4) {};
			\node[very small black vertex] at (0.5,5) {};
			\node[very small black vertex] at (1.5,5) {};
			\node[very small black vertex] at (1,5.5) {};
			\node[very small black vertex] at (1,4.5) {};
		\end{scope}
        
		\begin{scope}[xshift=-5 cm, yshift=0 cm]
            \draw[very thick] (0.75,1.25) -- (1.25,0.75);
            \draw[very thick, color=Red] (1,2) -- (0.75,1.25) (1,0) -- (1.25,0.75);
            \draw[very thick, color=RoyalBlue] (0.75,1.25) -- (0,1) (1.25,0.75)-- (2,1);

			\node[small black vertex] at (0,1) [label=left:$P_v$] {};
			\node[small black vertex] at (2,1) {};
			\node[small black vertex] at (1,2) [label=above:$Q_u$] {};
			\node[small black vertex] at (1,0) {};
			\node[very small black vertex] at (0.75,1.25) {};
			\node[very small black vertex] at (1.25,0.75) {};
		\end{scope}

		\draw[color=gray] (0,0) rectangle (6,6);
		
		\foreach \x in {0, 2}{
			\begin{scope}[shift={(\x cm, -\x cm)}]
				\draw[very thick, color=Red] (1,6) -- (1,4);
                \draw[very thick, color=RoyalBlue] (0,5) -- (2,5);
				\node[small black vertex] at (0.5,5) {};
				\node[small black vertex] at (1.5,5) {};
				\node[small black vertex] at (1,5.5) {};
				\node[small black vertex] at (1,4.5) {};
			\end{scope}
		}
        \foreach \x / \y in {2/0, 4/-2, 2/-4}{
            \begin{scope}[shift={(\x cm, \y cm)}]
				\draw[very thick, color=Red] (1,6) -- (1,4);
                \draw[very thick, color=RoyalBlue] (0,5) -- (2,5);
				\node[small black vertex] at (0.5,5) {};
				\node[small black vertex] at (1.5,5) {};
				\node[small black vertex] at (1,5.5) {};
				\node[small black vertex] at (1,4.5) {};
			\end{scope}

        }
		\foreach \x / \y in {0/0,0/2, 4/4, 4/0}{
		\begin{scope}[xshift=\x cm, yshift=\y cm]
            \draw[very thick] (0.75,1.25) -- (1.25,0.75);
            \draw[very thick, color=Red] (1,2) -- (0.75,1.25) (1,0) -- (1.25,0.75);
            \draw[very thick, color=RoyalBlue] (0.75,1.25) -- (0,1) (1.25,0.75)-- (2,1);
			\node[small black vertex] at (0.75,1.25) {};
			\node[small black vertex] at (1.25,0.75) {};
		\end{scope}
		}
        \foreach \i in {1,3,5}{
            \draw[dotted edge] (\i, 0) -- (\i,-1) (6,\i)--(7,\i);
			\draw[dotted edge] (\i, 6) -- (\i, 7) (-1, \i)--(0,\i);
			\foreach \j in {0,2,4,6}{
				\node[black vertex] (q\i\j) at (\i,\j) {};
				\node[black vertex] (p\j\i) at (\j,\i) {}; 
			}
		}

%% file: figures/red-from-mcc-example1.tex
		\node[round] (r1) at (0,0) [label=below:1] {};
		\node[round] (r2) at (1,0) [label=below:2] {};
		\node[square] (b1) at (-1,2) [label=left:1] {};
		\node[square] (b2) at (-1,1) [label=left:2] {};
		\node[lozenge] (g1) at (2,2) [label=right:1] {};
		\node[lozenge] (g2) at (2,1) [label=right:2] {};
		\node[triangle] (d1) at (0,3) [label=above:1] {};
		\node[triangle] (d2) at (1,3) [label=above:2] {};

		\draw[gray edge] (r2)--(g2)--(b2);
		\draw[gray edge] (r1) -- (b2);
		\draw[black edge, very thick] (r1)--(b1)--(g1)--(r1);
		\foreach \c in {r,g,b}{
			\draw[black edge, very thick] (\c1)--(d1);
			\draw[gray edge] (\c2)--(d1);
		}

%% file: figures/red-from-mcc-example2.tex
		\node[round] (s1) at (0,19) {};
		\node[round] (t1) at (23,19) {};
		\node[round] (s5) at (4,23) {};
		\node[round] (t5) at (4,0) {};
		\node[square] (s2) at (0,14) {};
		\node[square] (t2) at (23,14) {};
		\node[square] (s6) at (9,23) {};
		\node[square] (t6) at (9,0) {};
		\node[lozenge] (s3) at (0,9) {};
		\node[lozenge] (t3) at (23,9) {};
		\node[lozenge] (s7) at (14,23) {};
		\node[lozenge] (t7) at (14,0) {};
		\node[triangle] (s4) at (0,4) {};
		\node[triangle] (t4) at (23,4) {};
		\node[triangle] (s8) at (19,23) {};
		\node[triangle] (t8) at (19,0) {};
        \begin{scope}[on background layer]
        \foreach \yshift in {0, 5, 10, 15}{
            \begin{scope}[yshift=\yshift cm]
                \draw[line width=5pt, color=lightgray, opacity=0.5, cap=round] (0,4)--(2,5)--(21,5)--(23,4);
            \end{scope}
        }
        \foreach \xshift in {0,5,10,15}{
            \begin{scope}[xshift=\xshift cm]
                \draw[line width=5pt, color=lightgray, opacity=0.5, cap=round] (4,0)--(3,2)--(3,21)--(4,23);
            \end{scope}
        }
        \end{scope}

		\foreach \xshift / \yshift in {2/17, 7/12, 12/7, 17/2, 2/7, 12/17, 7/7, 12/12}{
		\begin{scope}[xshift=\xshift cm, yshift=\yshift cm]
			\draw[color=gray] (0,0) rectangle (4,4);
			\foreach \y in {1,3}{
				\foreach \x in {0,2,4}{
					\node[very small black vertex] at (\x, \y) {};
				}
			}
			\foreach \x in {1,3}{
				\foreach \y in {0,2,4}{
					\node[very small black vertex] at (\x, \y) {};
				}
			}
			\node[very small black vertex] at (0.5,3) {};
			\node[very small black vertex] at (1.5,3) {};
			\node[very small black vertex] at (1,2.5) {};
			\node[very small black vertex] at (1,3.5) {};
			\draw (0,3)--(2,3) (1,4)--(1,2);
			\begin{scope}[xshift=2cm, yshift=-2cm]
				\node[very small black vertex] at (0.5,3) {};
				\node[very small black vertex] at (1.5,3) {};
				\node[very small black vertex] at (1,2.5) {};
				\node[very small black vertex] at (1,3.5) {};
				\draw (0,3)--(2,3) (1,4)--(1,2);
			\end{scope}
			\node[very small black vertex] at (0.75,1.25) {};
			\node[very small black vertex] at (1.25,0.75) {};
			\draw (0,1)--(0.75,1.25)--(1,2);
			\draw (1,0)--(1.25,0.75)--(2,1);
			\draw (0.75,1.25)--(1.25,0.75);
			\begin{scope}[xshift=2cm, yshift=2cm]
				\node[very small black vertex] at (0.75,1.25) {};
				\node[very small black vertex] at (1.25,0.75) {};
				\draw (0,1)--(0.75,1.25)--(1,2);
				\draw (1,0)--(1.25,0.75)--(2,1);
				\draw (0.75,1.25)--(1.25,0.75);
			\end{scope}
		\end{scope}
		}
		\foreach \xshift / \yshift in {2/12}{
		\begin{scope}[xshift=\xshift cm, yshift=\yshift cm]
			\draw[color=gray] (0,0) rectangle (4,4);
			\foreach \y in {1,3}{
				\foreach \x in {0,2,4}{
					\node[very small black vertex] at (\x, \y) {};
				}
			}
			\foreach \x in {1,3}{
				\foreach \y in {0,2,4}{
					\node[very small black vertex] at (\x, \y) {};
				}
			}
			\node[very small black vertex] at (0.5,1) {};
			\node[very small black vertex] at (1.5,1) {};
			\node[very small black vertex] at (1,0.5) {};
			\node[very small black vertex] at (1,1.5) {};
			\draw (0,1)--(2,1) (1,2)--(1,0);
			\begin{scope}[yshift=2cm]
			\node[very small black vertex] at (0.5,1) {};
			\node[very small black vertex] at (1.5,1) {};
			\node[very small black vertex] at (1,0.5) {};
			\node[very small black vertex] at (1,1.5) {};
			\draw (0,1)--(2,1) (1,2)--(1,0);
			\end{scope}
			\node[very small black vertex] at (2.75,1.25) {};
			\node[very small black vertex] at (3.25,0.75) {};
			\draw (2,1)--(2.75,1.25)--(3,2);
			\draw (3,0)--(3.25,0.75)--(4,1);
			\draw (2.75,1.25)--(3.25,0.75);
			\begin{scope}[yshift=2cm]
			\node[very small black vertex] at (2.75,1.25) {};
			\node[very small black vertex] at (3.25,0.75) {};
			\draw (2,1)--(2.75,1.25)--(3,2);
			\draw (3,0)--(3.25,0.75)--(4,1);
			\draw (2.75,1.25)--(3.25,0.75);
			\end{scope}
			\end{scope}
		}
		\foreach \xshift / \yshift in {2/2, 7/17, 7/2, 12/2}{
		\begin{scope}[xshift=\xshift cm, yshift=\yshift cm]
		\draw[color=gray] (0,0) rectangle (4,4);
		\foreach \y in {1,3}{
		\foreach \x in {0,2,4}{
		\node[very small black vertex] at (\x, \y) {};
		}
		}
		\foreach \x in {1,3}{
		\foreach \y in {0,2,4}{
		\node[very small black vertex] at (\x, \y) {};
		}
		}
		\node[very small black vertex] at (0.5,3) {};
		\node[very small black vertex] at (1.5,3) {};
		\node[very small black vertex] at (1,2.5) {};
		\node[very small black vertex] at (1,3.5) {};
		\draw (0,3)--(2,3) (1,4)--(1,2);
		\begin{scope}[xshift=2cm]
		\node[very small black vertex] at (0.5,3) {};
		\node[very small black vertex] at (1.5,3) {};
		\node[very small black vertex] at (1,2.5) {};
		\node[very small black vertex] at (1,3.5) {};
		\draw (0,3)--(2,3) (1,4)--(1,2);
		\end{scope}
		\node[very small black vertex] at (0.75,1.25) {};
		\node[very small black vertex] at (1.25,0.75) {};
		\draw (0,1)--(0.75,1.25)--(1,2);
		\draw (1,0)--(1.25,0.75)--(2,1);
		\draw (0.75,1.25)--(1.25,0.75);
		\begin{scope}[xshift=2cm]
		\node[very small black vertex] at (0.75,1.25) {};
		\node[very small black vertex] at (1.25,0.75) {};
		\draw (0,1)--(0.75,1.25)--(1,2);
		\draw (1,0)--(1.25,0.75)--(2,1);
		\draw (0.75,1.25)--(1.25,0.75);
		\end{scope}
		\end{scope}
		}
		\foreach \xshift / \yshift in {17/17, 17/12, 17/7}{
		\begin{scope}[xshift=\xshift cm, yshift=\yshift cm]
			\draw[color=gray] (0,0) rectangle (4,4);
			\foreach \y in {1,3}{
		\foreach \x in {0,2,4}{
		\node[very small black vertex] at (\x, \y) {};
		}
		}
		\foreach \x in {1,3}{
		\foreach \y in {0,2,4}{
		\node[very small black vertex] at (\x, \y) {};
		}
		}
		\node[very small black vertex] at (0.5,3) {};
		\node[very small black vertex] at (1.5,3) {};
		\node[very small black vertex] at (1,2.5) {};
		\node[very small black vertex] at (1,3.5) {};
		\draw (0,3)--(2,3) (1,4)--(1,2);
		\begin{scope}[yshift=-2cm]
		\node[very small black vertex] at (0.5,3) {};
		\node[very small black vertex] at (1.5,3) {};
		\node[very small black vertex] at (1,2.5) {};
		\node[very small black vertex] at (1,3.5) {};
		\draw (0,3)--(2,3) (1,4)--(1,2);
		\end{scope}
		\node[very small black vertex] at (2.75,1.25) {};
		\node[very small black vertex] at (3.25,0.75) {};
		\draw (2,1)--(2.75,1.25)--(3,2);
		\draw (3,0)--(3.25,0.75)--(4,1);
		\draw (2.75,1.25)--(3.25,0.75);
		\begin{scope}[yshift=2cm]
		\node[very small black vertex] at (2.75,1.25) {};
		\node[very small black vertex] at (3.25,0.75) {};
		\draw (2,1)--(2.75,1.25)--(3,2);
		\draw (3,0)--(3.25,0.75)--(4,1);
		\draw (2.75,1.25)--(3.25,0.75);
		\end{scope}
		\end{scope}
		}
		\begin{scope}[on background layer]
		\foreach \yshift in {4, 9, 14, 19}{
		\begin{scope}[yshift=\yshift cm]
		\draw[dashed edge, gray] (0,0)--(2,-1);
		\draw[dashed edge, gray] (0,0)--(2,1);
		\draw[dashed edge, gray] (23,0)--(21,-1);
		\draw[dashed edge, gray] (23,0)--(21,1);
		\end{scope}
		}
		\foreach \xshift in {4, 9, 14, 19}{
		\begin{scope}[xshift=\xshift cm]
		\draw[dashed edge, gray] (0,0)--(-1,2);
		\draw[dashed edge, gray] (0,0)--(1,2);
		\draw[dashed edge, gray] (0,23)--(-1,21);
		\draw[dashed edge, gray] (0,23)--(1,21);
		\end{scope}
		}
		\foreach \xshift in {6, 11, 16}{
		\foreach \yshift in {3, 8, 13, 18} {
		\begin{scope}[xshift=\xshift cm, yshift=\yshift cm]
		\draw[dotted edge, gray] (0,0)--(1,0);
		\draw[dotted edge, gray] (0,2)--(1,2);
		\end{scope}
		\begin{scope}[xshift=\yshift cm, yshift=\xshift cm]
		\draw[dotted edge, gray] (0,0)--(0,1);
		\draw[dotted edge, gray] (2,0)--(2,1);
		\end{scope}
		}
		}
		\end{scope}
		\begin{scope}[on background layer]
		\draw[gray edge] (s5) to[bend left=10] node [very small gray vertex, pos=0.33] (v1) {} node [very small gray vertex, pos=0.66] (v2) {} (s6);
		\draw[gray edge] (s1) to[bend right=10] node [very small gray vertex, pos=0.33] {} node [very small gray vertex, pos=0.66] {} (s2);
		\draw[gray edge] (t1) to[bend left=10] node [very small gray vertex, pos=0.33] {} node [very small gray vertex, pos=0.66] {} (t2);
		\draw[gray edge] (s1) to node [very small gray vertex, pos=0.4] {} node [very small gray vertex, pos=0.6] {} (t1);
		\draw[gray edge] (s5) to node [very small gray vertex, pos=0.4] {} node [very small gray vertex, pos=0.6] {} (t5);
		\end{scope}

%% file: figures/red-from-3ec.tex
        \draw[color=lightgray] (-0.75,-0.75) rectangle (3.25,2.75);
        \node[draw=none] at (1.3,-1.1) {A};
        \node[black vertex] (1) at (0,2) [label=left:1] {};
        \node[black vertex] (2) at (0,0) [label=left:2] {};
        \node[black vertex] (3) at (1.5,1) [label=above:3] {};
        \node[black vertex] (4) at (3,1) [label=above:4] {};
        \begin{scope}[on background layer]
            \draw[dashed edge, red] (1)--(2);
            \draw[dotted edge, Green] (1)--(3);
            \draw[dashdotted edge, RoyalBlue] (2)--(3);
            \draw[dashed edge, red] (3)--(4);
        \end{scope}

        \begin{scope}[xshift=3cm]
        \draw[color=lightgray] (0.5,-0.75) rectangle (4.5,2.75);
        \node[draw=none] at (2.6,-1.1) {B};
            \foreach [count=\i] \color in {r,g,b,d} {
                \node [black vertex] (x\color) at (\i, 2) [label=above:$x_{\color}$]{};
                \node[black vertex] (v\i) at (\i,0) [label=below:$\i$] {};}
            \foreach \i in {1,...,4}{
                \foreach \color in {r,g,b,d}{
                    \draw[gray edge] (v\i) -- (x\color);
                }
            }
            \draw[dashed arc, red] (v1) to (v2);
            \draw[dotted arc, Green] (v1) to[bend right=30] (v3);
            \draw[dashdotted arc, RoyalBlue] (v2) to (v3);
            \draw[dashed arc, red] (v3) to (v4);
            \foreach \i in {1,...,4}{
                \draw[black arc] (xd) -- (v\i);}
        \end{scope}

        \begin{scope}[xshift=7.25cm]
        \draw[color=lightgray] (0.5,-0.75) rectangle (4.5,2.75);
        \node[draw=none] at (2.6,-1.1) {C};
            \foreach [count=\i] \color in {r,g,b,d}{
                \node[black vertex] (y\color) at (\i, 2) [label=above:$x_{\color}$]{};
                \node[black vertex] (u\i) at (\i,0) [label=below:$\i$] {};}
            \foreach \i in {1,...,4}{
            \draw[black edge] (yd)--(u\i);};
            \draw[dashed edge, red] (u1)--(yr)--(u2);
            \draw[dotted edge, Green] (u1)--(yg)--(u3);
            \draw[dashdotted edge, RoyalBlue] (u2)--(yb)--(u3);
            \draw[dashed edge, red] (u3)--(yr)--(u4);
            \draw[gray edge] (u2)--(yg)--(u4);
            \draw[gray edge] (u1)--(yb)--(u4);
        \end{scope}

%% file: figures/red-from-bp.tex
			\foreach \i in {1,2,3}{
				\begin{scope}[xshift=(\i-1)*1.5cm]
					\draw[fill=white] (0,0,0)--(0.75,0,0)--(0.75,0.75,0)--(0,0.75,0)--cycle;
					\draw[fill=white] (0,0,0)--(0,0,0.75)--(0,0.75,0.75)--(0,0.75,0)--cycle;
					\draw[fill=white] (0,0,0.75)--(0.75,0,0.75)--(0.75,0.75,0.75)--(0,0.75,0.75)--cycle;
					\draw[fill=white] (0.75,0.75,0)--(0.75,0,0)--(0.75,0,0.75)--(0.75,0.75,0.75)--cycle;
					\node at (0.1,0.1) {$B_{\i}$}; 
				\end{scope}
			}
			\node[draw, fill=lightgray] (2) at (0.25,2) {2};
			\node[draw, fill=lightgray] (3) at (1.25,2) {3};
			\node[draw, fill=lightgray] (4) at (2.25,2) {4};
			\node[draw, fill=lightgray] (5) at (3.25,2) {5};
			\draw[->, very thick] (2)--(0.25,0.55);
			\draw[dotted arc, Green] (3)--(0.4,0.55);
			\draw[dashed arc, red] (4)--(1.8,0.55);
			\draw[dashdotted arc, RoyalBlue] (5)--(3.25,0.55);

			\begin{scope}[xshift=5cm, yshift=1cm]
				\node[black vertex] (s) at (0, 0) [label=left:$s$] {};
				\node[black vertex] (t) at (4, 0) [label=right:$t$] {};
				\node[draw] (B1) at (2,0.75) {$B_1$};
				\node[draw] (B2) at (2,0) {$B_2$};
				\node[draw] (B3) at (2,-0.75) {$B_3$};
				\draw[very thick] (s)--(B1)--(t);
				\draw[dotted edge, Green] (s) to[bend left=15] (B1) ;
				\draw[dotted edge, Green] (B1) to[bend left=15] (t);
				\draw[dashed edge, red] (s)--(B2)--(t);
				\draw[dashdotted edge, RoyalBlue] (s)--(B3)--(t);
				\draw[gray edge] (s) to[bend right=90] (t) ;
				\end{scope}

%% file: figures/cycles.tex
        \begin{scope}[xshift=0cm]
            \node[small black vertex] (a1) at (0,0) {};
            \node[small black vertex] (a2) at (-0.5,0.25) [label=left:$s_2$] {};
            \node[small black vertex] (a3) at (-0.5,0.75) [label=left:$s_1$] {};
            \node[small black vertex] (a4) at (0,1) {};
            \node[small black vertex] (a5) at (0.5,0.75) [label=right:$t_2$] {};
            \node[small black vertex] (a6) at (0.5,0.25) [label=right:$t_1$] {};
            \draw (a1)--(a2)--(a3)--(a4)--(a5)--(a6)--(a1);
        \end{scope}
        \begin{scope}[xshift=3cm]
            \node[small black vertex] (b1) at (0,0) {};
            \node[small black vertex] (b2) at (-0.5,0.25) {};
            \node[square waypoint] at (b2) {};
            \node[small black vertex] (b3) at (-0.5,0.75) {};
            \node[square waypoint] at (b3) {};
            \node[small black vertex] (b4) at (0,1) {};
            \node[small black vertex] (b5) at (0.5,0.75) [label=right:$s_1 t_2$] {};
            \node[small black vertex] (b6) at (0.5,0.25) [label=right:$t_1 s_2$] {};
            \draw (b1)--(b2)--(b3)--(b4)--(b5)--(b6)--(b1);
        \end{scope}
        \begin{scope}[xshift=6cm]
            \node[small black vertex] (c1) at (0,0) [label=below:$t_1 t_2$] {};
            \node[small black vertex] (c2) at (-0.5,0.25) {};
            \node[small black vertex] (c3) at (-0.5,0.75) {};
            \node[small black vertex] (c4) at (0,1) [label=above:$s_1 s_2$] {};
            \node[small black vertex] (c5) at (0.5,0.75) [label=right:$s_3$] {};
            \node[small black vertex] (c6) at (0.5,0.25) [label=right:$t_3$] {};
            \draw (c1)--(c2)--(c3)--(c4)--(c5)--(c6)--(c1);
        \end{scope}

%% file: figures/cactus-skeleton.tex
        \node[small black vertex] (c11) at (-0.5,0.5) {};
        \node[small black vertex] (c12) at (-1,0.25) {};
        \node[small black vertex] (c13) at (-1,-0.25) {};
        \node[small black vertex] (c14) at (-0.5,-0.5) {};
        \node[draw, inner sep=1pt] (h4) at (0,0) {$h_4$};
        \draw (c11)--(c12)--(c13)--(c14)--(h4)--(c11);

        \node[small black vertex] (g21) at (0,0.75) {};
        \draw (g21)--(h4);

        \node[small black vertex] (g31) at (0, -0.75) {};
        \node[small black vertex] (g32) at (-0.25, -1) {};
        \node[small black vertex] (g33) at (0.25, -1) {};
        \draw (h4)--(g31)--(g32) (g31)--(g33);
        \node[draw, inner sep=1pt] (h2) at (1,0) {$h_2$};
        \draw (h4)--(h2);

        \node[small black vertex] (g41) at (1, 0.5) {};
        \node[draw, inner sep=1pt] (h1) at (0.75, 1.25) {$h_1$};
        \node[draw, inner sep=1pt] (h3) at (1.25, 1.25) {$h_3$};
        \draw (h2)--(g41)--(h1) (g41)--(h3);

        \node[small black vertex] (c21) at (0.5,2.25) {};
        \node[small black vertex] (c22) at (-0.25,2.25) {};
        \node[small black vertex] (c23) at (-0.25,1.5) {};
        \draw (h1)--(c21)--(c22)--(c23)--(h1);

        \node[small black vertex] (c31) at (2,2.25) {};
        \node[small black vertex] (c32) at (2.25,1.5) {};
        \draw (h3)--(c31)--(c32)--(h3);

        \node[small black vertex] (c41) at (1.65,0.5) {};
        \node[draw, inner sep=1pt] (h6) at (1.65,-0.25) {$h_6$};
        \node[draw, inner sep=1pt] (h5) at (2.25,0) {$h_5$};
        \draw (h2)--(c41)--(h5)--(h6)--(h2);

        \node[small black vertex] (g11) at (2.25,0.5) {};
        \node[small black vertex] (g12) at (2,1) {};
        \node[small black vertex] (g13) at (2.5,1) {};
        \draw (h5)--(g11)--(g12) (g11)--(g13);
        
        \node[small black vertex] (c51) at (1.5,-1) {};
        \node[small black vertex] (c52) at (2.1,-1) {};
        \draw (h6)--(c51)--(c52)--(h6);
        
        \node (C1) at (-0.6,0) {$c_1$};
        \node (C2) at (0.1,1.9) {$c_2$};
        \node (C3) at (1.9,1.75) {$c_3$};
        \node (C4) at (1.65,0.15) {$c_4$};
        \node (C5) at (1.75,-0.75) {$c_5$};
        \begin{scope}[on background layer]
            \draw[dashed, darkgray] (0,-0.5) ellipse (0.4cm and 0.7cm);
            \draw[dashed, darkgray] (0,0.3) ellipse (0.3cm and 0.6cm);
            \draw[dashed, darkgray] (0.5,0) ellipse (0.7cm and 0.3cm);
            \draw[dashed, darkgray] (1,0.8) ellipse (0.5cm and 1cm);
            \draw[dashed, darkgray] (2.25,0.5) ellipse (0.4cm and 0.7cm);
        \end{scope}
        \node (G1) at (2.25,1) {$g_1$};
        \node (G2) at (-0.15,0.5) {$g_2$};
        \node (G3) at (-0.15,-0.5) {$g_3$};
        \node (G4) at (1.25,0.5) {$g_4$};
        \node (G5) at (0.55,-0.2) {$g_5$};

        \begin{scope}[xshift=5cm, yshift=2cm]
            \node[draw, circle, inner sep=1pt] (g4) at (0,0) {$g_4$};
            \node[draw, inner sep=1pt] (H1) at (-1,-0.75) {$h_1$};
            \node[draw, inner sep=1pt] (H2) at (0,-0.75) {$h_2$};
            \node[draw, inner sep=1pt] (H3) at (1,-0.75) {$h_3$};
            \node[draw, circle, inner sep=1pt] (c2) at (-1.5,-1.5) {$c_2$};
            \node[draw, circle, inner sep=1pt] (c4) at (-0.5,-1.5) {$c_4$};
            \node[draw, circle, inner sep=1pt] (g5) at (0.5,-1.5) {$g_5$};
            \node[draw, circle, inner sep=1pt] (c3) at (1.5,-1.5) {$c_3$};
            \node[draw, inner sep=1pt] (H6) at (-1.25,-2.25) {$h_6$};
            \node[draw, inner sep=1pt] (H5) at (-0.5,-2.25) {$h_5$};
            \node[draw, inner sep=1pt] (H4) at (0.75,-2.25) {$h_4$};
            \node[draw, circle, inner sep=1pt] (g1) at (-0.5,-3) {$g_1$};
            \node[draw, circle, inner sep=1pt] (c5) at (-1.25,-3) {$c_5$};
            \node[draw, circle, inner sep=1pt] (c1) at (0.25,-3) {$c_1$};
            \node[draw, circle, inner sep=1pt] (g2) at (1,-3) {$g_2$};
            \node[draw, circle, inner sep=1pt] (g3) at (1.75,-3) {$g_3$};
            \draw (g4)--(H1) (g4)--(H2) (g4)--(H3);
            \draw (c2)--(H1) (c4)--(H2)--(g5) (c3)--(H3);
            \draw (H6)--(c4)--(H5)--(g1) (g5)--(H4);
            \draw (H6)--(c5) (c1)--(H4)--(g2) (H4)--(g3);
        \end{scope}

%% file: figures/red-from-3partition.tex
        \foreach [count=\i] \xshift in {0,2}{
        \begin{scope}[xshift=\xshift cm]
            \node[gray vertex] (ta\i) at (0,1) {};
            \node[gray vertex] (tb\i) at (0.5,0) {};
            \node[gray vertex] (tc\i) at (1,1) {};
            \draw[gray edge] (ta\i)--(tb\i)--(tc\i)--(ta\i);
            \node at (0.5,0.65) {$T_\i$};
        \end{scope}
        }
        \draw[gray edge] (tc1)--(ta2);
        \node[black vertex] at (ta1) [label=left:$s$] {};
        \node[black vertex] at (tc2) [label=right:$t$] {};
        \node[square waypoint] at (tb1) [label=below:{$\lbrace 4,5,6 \rbrace$}] {};
        \node[round waypoint] at (tb2) [label=below:{$\lbrace 5,5,5 \rbrace$}] {};